\documentclass[12pt]{article}
\usepackage{graphicx} 
\usepackage{subcaption}
\usepackage{booktabs}
\usepackage{float}
\usepackage{setspace}
\usepackage[a4paper, margin=1in]{geometry}
\usepackage{amsmath}
\usepackage[utf8]{inputenc}
\newcommand{\ThesisTitle}{A DC discharge plasma \\\vspace{0.4cm} experiment for undergraduate laboratories}
\newcommand{\YourName}{You-Hsuan Chen\textsuperscript{1,2*}, Ting-An Wang\textsuperscript{1,2*}, and Pisin Chen\textsuperscript{1,2}}

\newcommand{\AffilOne}{\textsuperscript{1}Department of Physics, National Taiwan University, Taipei 10617, Taiwan, ROC.}
\newcommand{\AffilTwo}{\textsuperscript{2}Leung Center for Cosmology and Particle Astrophysics, National Taiwan University, Taipei 10617, Taiwan, ROC.}

\begin{document}
\begin{titlepage}

\thispagestyle{empty}

\vspace{1.5cm}

\begin{center}

    {\LARGE \bfseries \ThesisTitle \par}
    \vspace{0.5cm}

    {\large \YourName}\\[0.2cm]

{\large
    \AffilOne \\[0.1cm]
    \AffilTwo \\[1.5cm]
    
}
\end{center}
\centerline{\textbf{Abstract}}\vspace{-0.5cm}
Plasma physics offers a wide range of fundamental phenomena, making it an excellent subject for undergraduate laboratory instruction. In this work, we present the design, construction, and characterization of a DC glow-discharge plasma chamber developed for the junior-level curriculum at National Taiwan University, a project carried out by two undergraduate students. The apparatus consists of a 1-meter-long quartz tube with a movable electrode, enabling systematic exploration of plasma behavior under varying pressure, voltage, and geometry. Using this platform, we characterized the Paschen breakdown relation and the voltage-current characteristics of the plasma. We then developed Langmuir probes to map spatial distributions of electron temperature and density, and used Boltzmann plot spectroscopy to measure excitation temperatures across different plasma regions. Finally, with custom Helmholtz coils, we demonstrated magnetic focusing of electrons. We performed Runge–Kutta simulations of particle trajectories and analyzed the electron drift velocity by comparing the focal lengths. Overall, this plasma chamber provides a versatile platform for investigating fundamental plasma phenomena and offers potential for future studies, including microwave–plasma interactions and other student-driven investigations.


\small{*} These authors contribute equally to this work\\
{Corresponding Author(s) and Email(s)}\\
Ting-An Wang: b11202003@ntu.edu.tw\\
You-Hsuan Chen:
b11202006@ntu.edu.tw

\vfill
\end{titlepage}
\tableofcontents
\clearpage
\doublespacing

\section{Introduction}\label{chap:intro}
Over 99\% of the baryonic matter in the universe exists in the plasma state, and consequently modern plasma physics spans several major research frontiers. The study of astrophysical plasmas provides insight into a wide range of cosmic phenomena, including stellar structure, astrophysical jets, and solar and magnetospheric winds \cite{Gekelman2016LAPD}. The pursuit of controlled fusion as a clean, sustainable energy source drives intensive research on magnetic-confinement devices such as tokamaks and stellarators \cite{RevModPhys.95.025005}. Plasma wakefield acceleration, driven by intense lasers \cite{PhysRevLett.43.267} or by electron beams \cite{Chen1985PWFA}, has enabled major breakthroughs for next-generation accelerators, achieving acceleration gradients three orders of magnitude higher than conventional RF cavities and thereby offering a promising pathway toward future high-energy colliders. The collective behavior of plasmas, governed by electromagnetic interactions, gives rise to dynamics that differ fundamentally from those in neutral gases or condensed matter. Understanding these behaviors is essential both for interpreting astrophysical observations and for advancing plasma-based technologies, tightly linking laboratory experiments, theoretical models, and the large-scale processes that operate throughout the universe.

Not only is plasma rich in physics, it is also a system that is relatively inexpensive to construct, making it an ideal system for undergraduates to study. Inspired by the work in Princeton Plasma Physics Laboratory \cite{Wissel2013}, we built a DC discharge plasma chamber as part of the "Fundamentals of Experimental Physics" junior lab course in the Department of Physics, National Taiwan University. We designed and built a DC discharge chamber, and performed a series of study on the behavior of DC discharge plasma. While \cite{Wissel2013} proposed studying many of these phenomena, we found that their work left room for further investigation, particularly when combined with spatial-distribution measurements and numerical simulations. Moreover, we designed our experiments such that all measurements could be performed within a single plasma chamber. Our plasma chamber setup will be part of the NTU Physics undergraduate curriculum starting in the fall semester of 2026. 

In section \ref{chap:intro}, we describe the motivation for building a plasma chamber and the related experiments from Princeton. In section \ref{chap:build} we discuss the design of the DC discharge plasma chamber and some plasma electrical characteristics, including breakdown voltage Paschen curve and current-voltage relations. This is followed by section \ref{chap:diag}, where we present the methods we developed to measure plasma temperature and density, including Langmuir probes and Boltzmann plots. Section \ref{chap:dyn} studies the dynamical properties of plasma, including the interaction of plasma with magnetic fields. Finally, section \ref{chap:conclusion} describes our current and future work with this apparatus, and conclusions of this experiment.

\subsection{Motivation}
The primary motivation for building a DC discharge plasma chamber was to create an apparatus suitable for the undergraduate physics teaching laboratory. While plasma physics plays a central role in many areas of science and technology, it is rarely included in the standard undergraduate curriculum. By providing a hands-on plasma experiment, students can directly observe and understand the underlying physics of ionized gases. Despite its simplicity, a DC discharge system embodies rich physical principles: optical spectroscopy reveals thermal equilibrium between atomic species and free electrons; applying magnetic fields demonstrates charge-to-mass-dependent motion and the separation of electrons and ions; and measuring the breakdown curve introduces concepts such as mean free path and electron-neutral collisions. Furthermore, the chamber is designed with a high degree of flexibility. The geometry, applied voltage, and gas pressure can all be adjusted. This allows students to explore a wide range of plasma behaviors while developing an intuitive understanding of this fundamental state of matter.

\subsection{Literature Review}
Our work is mostly inspired by the paper "The use of dc glow discharges as undergraduate educational tools"\cite{Wissel2013} by the Princeton Plasma Physics Laboratory. In the paper, they built several DC plasma devices that allow students to study:\\
\textbf{1. Electrical breakdown of Argon gas:} At various pressures, the voltage across the electrodes required to break down the background neutral gas is measured. The curve connecting the voltages across different pressures can be described by the Paschen theory by considering the balance of the rate of electron production and ion bombardment. Paschen's theory is described in detail in \ref{chap:Paschen}.\\
\textbf{2. Plasma voltage-current (V-I) characteristics:} At various pressures, the voltage across the plasma is varied and resulting discharge current is measured. The voltage-current characteristics differ across the distinct regimes of DC discharge, as shown in Fig.~\ref{'IV_ref'}.\\
\textbf{3. Plasma temperature and density measurement:} The standard way to perform plasma diagnostics is to develop a Langmuir probe and insert it into the plasma chamber. By biasing the probe relative to the background plasma potential and measuring the current picked up by the probe, the plasma (electron) density and temperature can be inferred. Other parameters, including ionization ratio, can be further deduced.\\
\textbf{4. Plasma emission spectroscopy:} DC discharge plasma emits light. The intensity of the transition lines from various atomic/molecular levels with the Einstein A coefficient can be used to deduce the population on each level, the process is called making a Boltzmann plot. From the slope of the Boltzmann plot, one can infer electronic temperature within the atom/molecule.\\
\textbf{5. Plasma--magnetic field interaction:}
Plasma is composed of an ionized gas containing electrons and ions. When a magnetic field is applied, the trajectories of charged particles are affected according to the Lorentz force. Because electrons are much lighter than ions, their paths are bent far more strongly. As a result, changes in the electron trajectory become visually apparent in the plasma structure and emission profile.

In this paper, we report our recreation of these five experiments. Compared with the work from Princeton, we have made several modifications:\\
a. All measurements are performed on the same chamber, making the setup more accessible as an undergraduate research project.\\
b. We conducted an in-depth study of the Langmuir probe, including different probe geometries, probe orientations, and the spatial dependence of plasma temperature and density. This enables a detailed structural analysis of DC discharge plasma.\\
c. We performed Boltzmann plot studies at different locations along the plasma column, observing spectral variations across the various regions.\\
d. We provided a detailed analysis of magnetic lensing and compare the results with simulation codes.\\
Our effort will make this experiment complete and allow students to characterize the system quantitatively.

\section{Building a DC Discharge Plasma Chamber}\label{chap:build}

\subsection{Producing Plasma by DC High Voltage }\label{chap:theory_of_dc}
One simple way of producing plasma is by applying a high DC voltage to some dilute gas, and the produced plasma is called a "DC discharge plasma".
The process of producing plasma proceeds as follows. Small, random ionization events caused by cosmic-ray collisions are always present within a neutral gas. As the voltage on the electrodes is increased, the resulting electric field accelerates electrons and ions in opposite directions. When the accelerated electrons acquire sufficient energy and collide with neutral atoms or molecules, they can ionize them, generating additional electrons and ions. This process can be described as:
\begin{equation}
e^-+A=2e^-+A^+
\end{equation}
This cascading ionization process is called the Townsend avalanche \cite{princeton_ref}. In this voltage regime, known as the Townsend Regime, the plasma is not stable and appears as a flickering glow. A key step in the transition to a glow discharge is secondary electron emission, which balances the number of electrons collected by the anode with those emitted by cathode. This happens when energetic ions collide with the cathode surface, forming free electrons that are stripped from the metal plates.
\begin{equation}
A^++M=e^-+A^++M^+
\end{equation}
The plasma is now in a self-sustainable state, appearing as a constant, visible glow \cite{smirnov2015theory}. The voltage required to initiate this glow discharge is known as the breakdown voltage. The regime examined in this report corresponds to the normal glow phase of a DC glow discharge.
\begin{figure}[H]
    \centering
    \includegraphics[width=0.75\columnwidth]{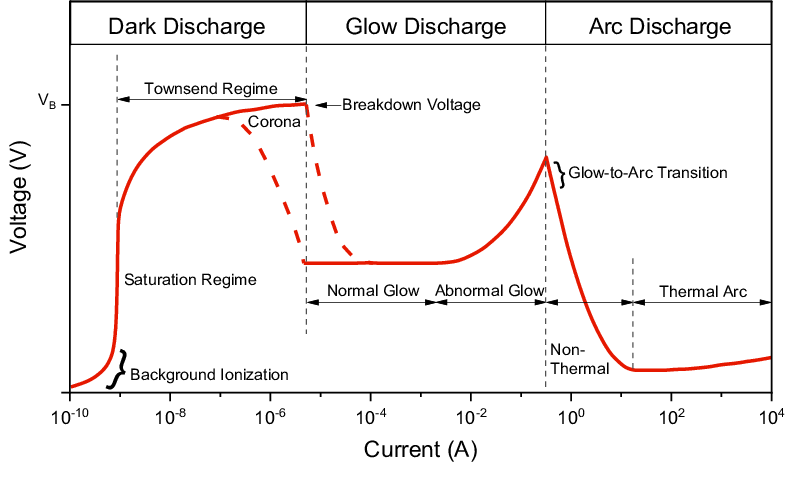}
    \caption{\textbf{Voltage-current curve of different stages of DC discharge. This graph is from \cite{IV_ref}}. The Normal Glow discharge is studied in this thesis. This regime exhibits Ohm's law relation between current and voltage. The regime of DC discharge can be driven by a low power high voltage DC power supply that is common in undergraduate  laboratories. }
    \label{'IV_ref'}
\end{figure}

In a DC glow discharge plasma, the emission is visibly inhomogeneous. One can characterize this behavior by identifying the distinct regions of the discharge.

\begin{figure}[H]
    \centering
    \includegraphics[width=1\columnwidth]{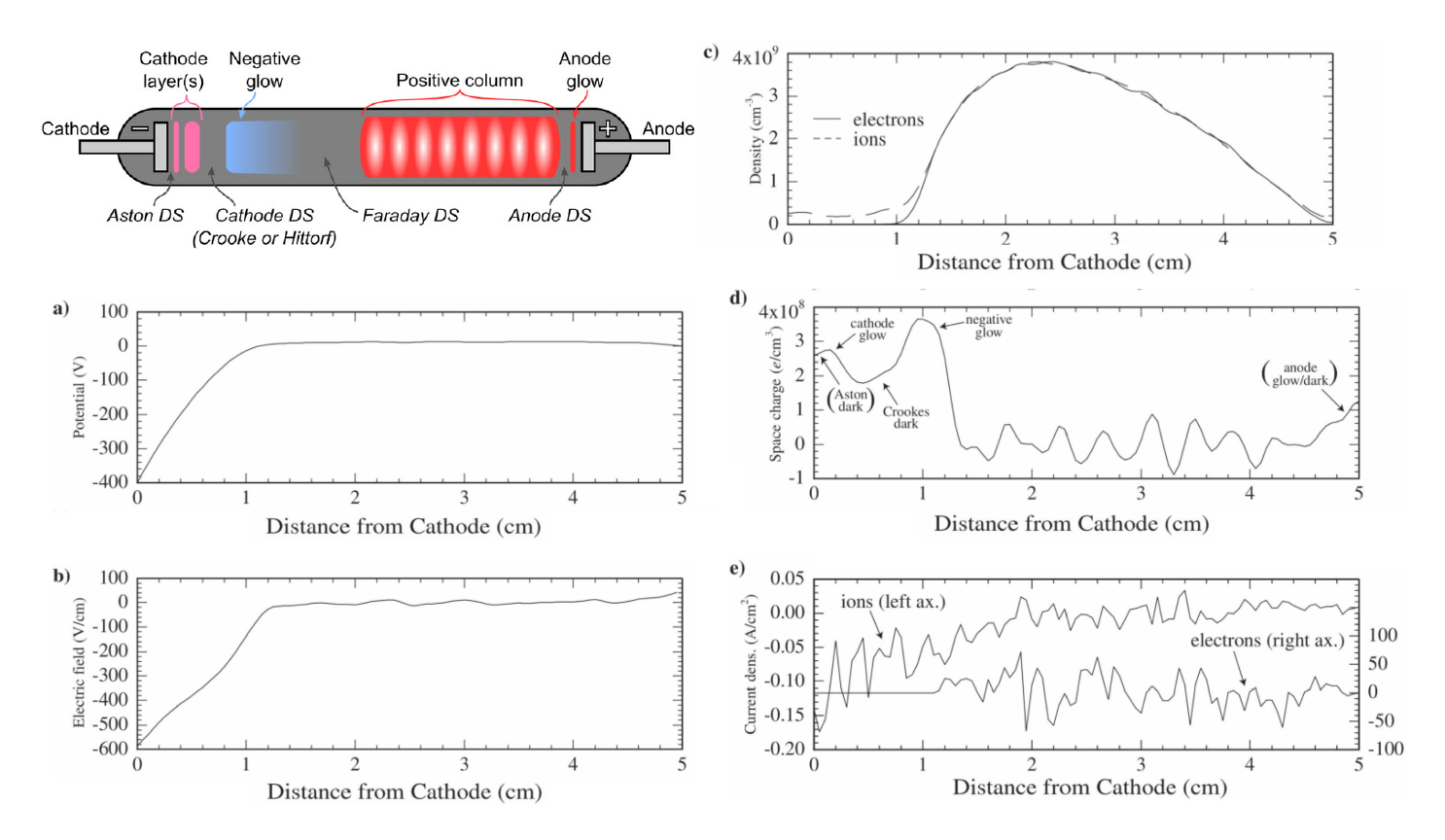}
    \caption{\textbf{Position dependent measurements in a DC glow discharge from \cite{fundamentals_of_dc} } This figure shows the plasma potential, plasma density, space charge, and current across the various regimes of a DC glow discharge.  }
    \label{fig:glows}
\end{figure}

Starting with the cathode, there is a space right next to the cathode with low luminance called the Aston dark space. Here, the secondary electrons emitted are accumulated. The electrons are relatively slow in this area as they are still in the process of acceleration. Next to the Aston dark space is the Cathode glow. This is where the electrons gain sufficient energy to ionize atoms through collision. Thus, the distance between the cathode glow and the cathode corresponds to the mean free path. Next to the cathode glow is the Crooke dark space. In this area, most space charge are positive. Even though there is frequent ionization, electrons and ions travel fast towards opposite directions, so there is rare recombination, giving off a low luminance appearance. The next region observed is the negative glow, where the electron velocity isn't as fast due to  the low electric field strength. Here, the intensity of emission is the highest across the chamber. Fig. \ref{fig:glows} shows the spatial variations of the potential, electric field, particle densities, space charge, and current densities along the axis. In (a) and (b), we can see that the electric field is at maximum near the cathode, and is almost zero starting at the negative glow. This implies the potential drop over the cathode sheath contributes to most of the potential difference. As we leave the cathode sheath, the electrons accelerate less due to the absence of electric field. The first region observed is then the Faraday dark space, where there is low ionization rate. After the Faraday dark space is the positive column. This is when the electrons gain enough energy from the low electric field, and start to ionize particles again, leading to the low luminous glow. Positive columns are quasi-neutral, except for the local inhomogeneity known as striations. 

However, the structures observed in the DC discharge varies with the gas species, pressure and applied voltage. Thus, the ionization and recombination rate across the chamber should be studied case by case.

\subsection{The Design of Our Plasma Chamber}
\begin{figure}[h!]
    \centering
    \includegraphics[width=\textwidth]{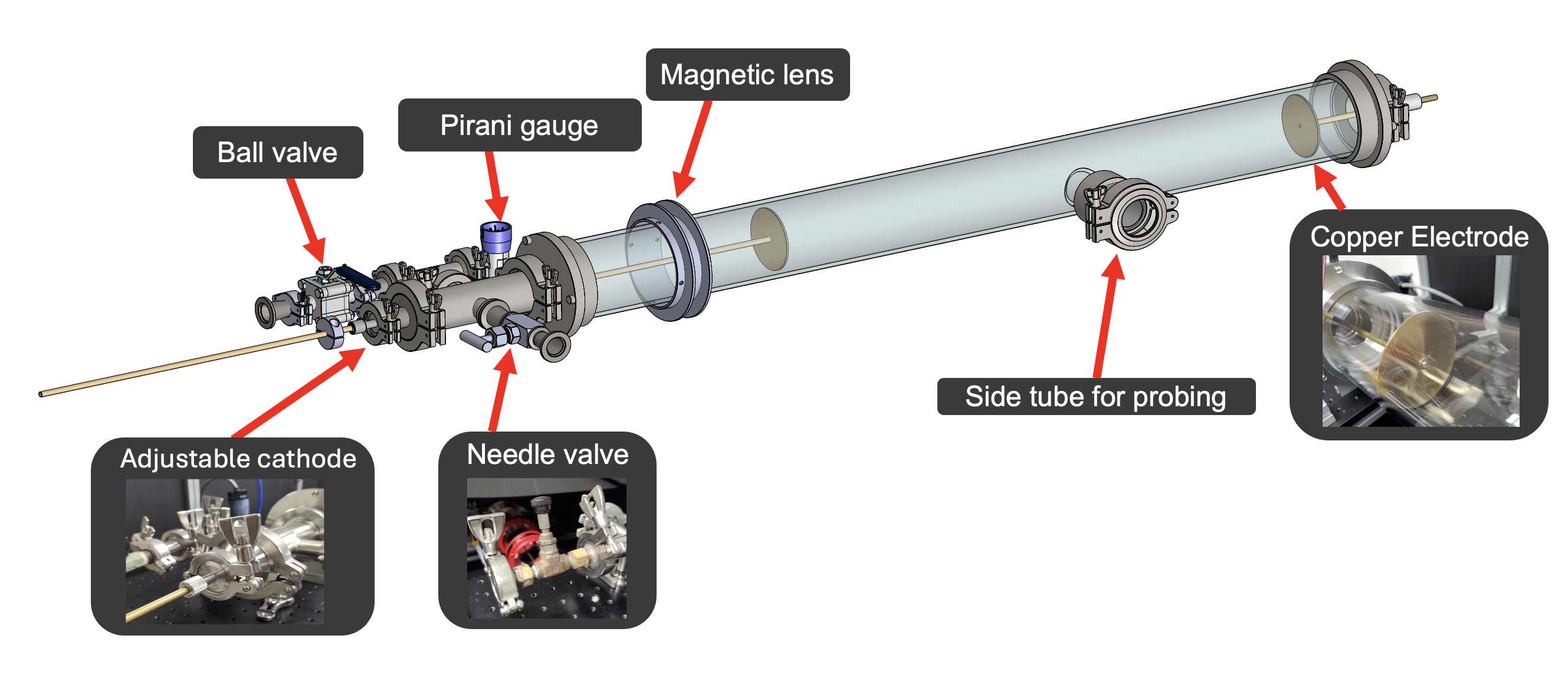}
    \caption{\textbf{DC discharge chamber design.}\\ 
    The main structure of the chamber is a 1.1-meter-long T-shape quartz glass cell. There are two sides of vacuum components connected to the chamber by O-rings, the left side is the cathode, the right anode. There is a side tube in the middle.\\
    1. Cathode section: we have an position-adjustable copper cathode that's sealed by compression fitting, a ball valve connected to the mechanical pump, a vacuum gauge meter, and a home-made needle valve in series with a ball valve to control the input flow of gas. This side is completely grounded.\\
    2. Anode section: A brass plate is screwed onto a oxygen-free copper HV feed through. The high voltage power supply is connected to this side of the chamber. During experiments, this side is shielded by a big 3D-printed cylinder to avoid electric shocks.\\
    3. The side-tube section: A KF50 quartz window installed for laser access and spectroscopy.}
    \label{fig:setup_exp}
\end{figure}

\begin{figure}[H]
    \centering
    \includegraphics[width=\textwidth]{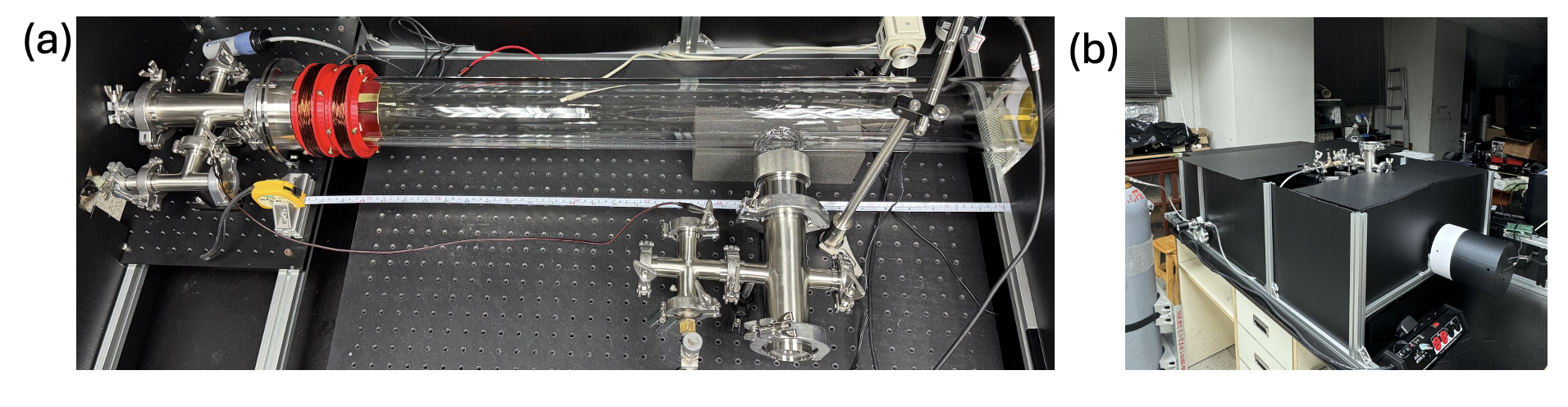}
    \caption{\textbf{Images of the chamber.} \\
    (a) The chamber without foam boards covering. The entire setup is supported by aluminum extrusions. The anode is covered with a plastic cylinder at the right, the cathode is at the left. In this image, a CCD camera is set pointing at the anode. (b) During the experiment, the chamber is covered up with black foam boards.}
    \label{fig:real_set}
\end{figure}

The design of the chamber is illustrated in Fig.~\ref{fig:setup_exp}. Plasma is generated between two copper electrodes in a 1.1-meter-long quartz glass cell. Quartz glass was chosen as the material for the chamber walls to minimize the optical absorption. A constant flow of working gas (helium) is maintained in the chamber, with the pressure preserved by a mechanical pump removing the excess gas. A needle valve and regulator are used to fine-tune the working gas flow rate. Typical pressures for our experiment is between 1 to 1000 mtorr. \\
Except for the anode, all vacuum components are grounded, including the cathode electrode. The high-voltage power supply we use can produce voltage up to 10kV with respect to ground, and is connected to the cathode. A special feature of our chamber is that the electrode distance is adjustable. The cathode is fed into the vacuum by the Thorlabs 1/4" compression fitting port. This allows us to study plasma formation at different electrode distances. \\
The chamber is supported by aluminum extrusions. Below the chamber is an optics table for setting up optical components such as CCDs and spectrometers. During measurements, the chamber is covered with black foam boards to avoid light contamination. A photo of the chamber is shown in Fig.~\ref{fig:real_set}.

\subsubsection{Electronic setup}
\begin{figure}[H]
    \centering
    \includegraphics[width=\textwidth]{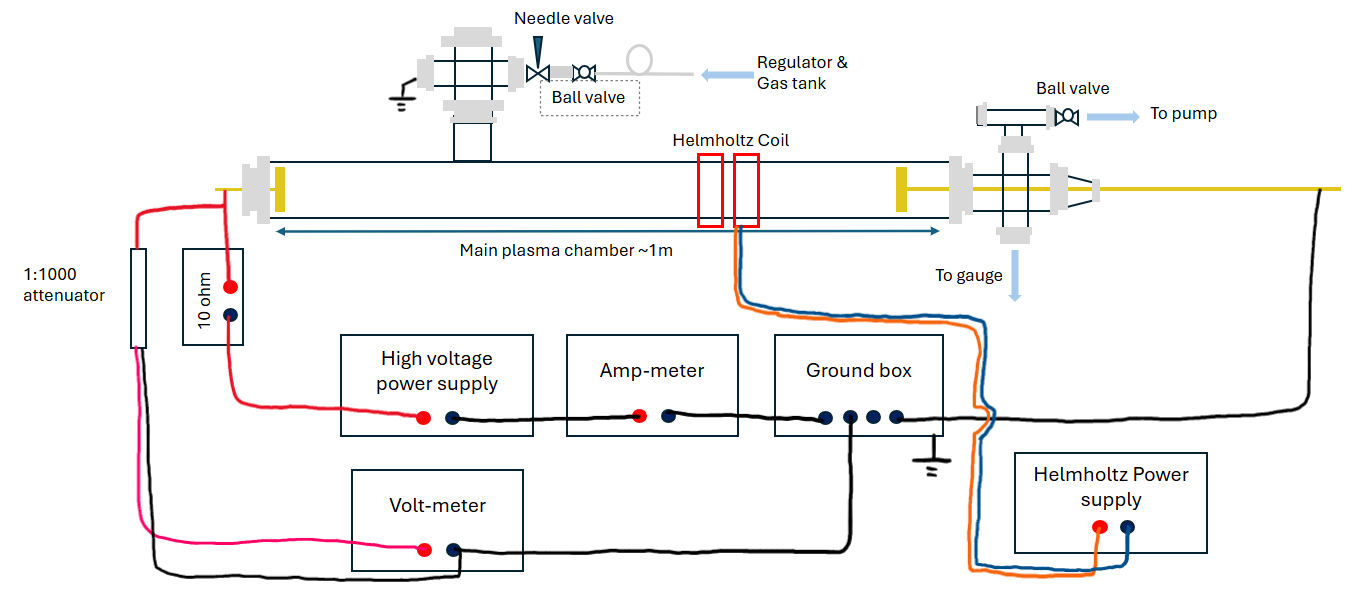}
    \caption{\textbf{Schematic diagram of the electronic layout of our plasma experiment.}\\
    The red wires correspond to high voltage (kV) cables, black correspond to ground. The high-voltage power supply creates kilovolts of potential difference relative to the ground. The current passes through a high voltage ceramic resistor, than connected to the anode. On the other side, the cathode is grounded. An amp meter is connected in series to monitor the current passing through the chamber. An individual 5A power supply provides current for the magnetic coils.} 
    \label{fig:circuit}
\end{figure}

The electronic setup is described in this section. See Fig.~\ref{fig:circuit} for the electronic layout. Our high-voltage power supply can provide voltage up to 10 kV with maximum power 20 W. At pressures between 50 to 700 mtorr, the breakdown voltage is between 0.8 to 2.0 kV.

\subsubsection{Graphical User Interface}
\begin{figure}[H]
    \centering
    \includegraphics[width=0.5\textwidth]{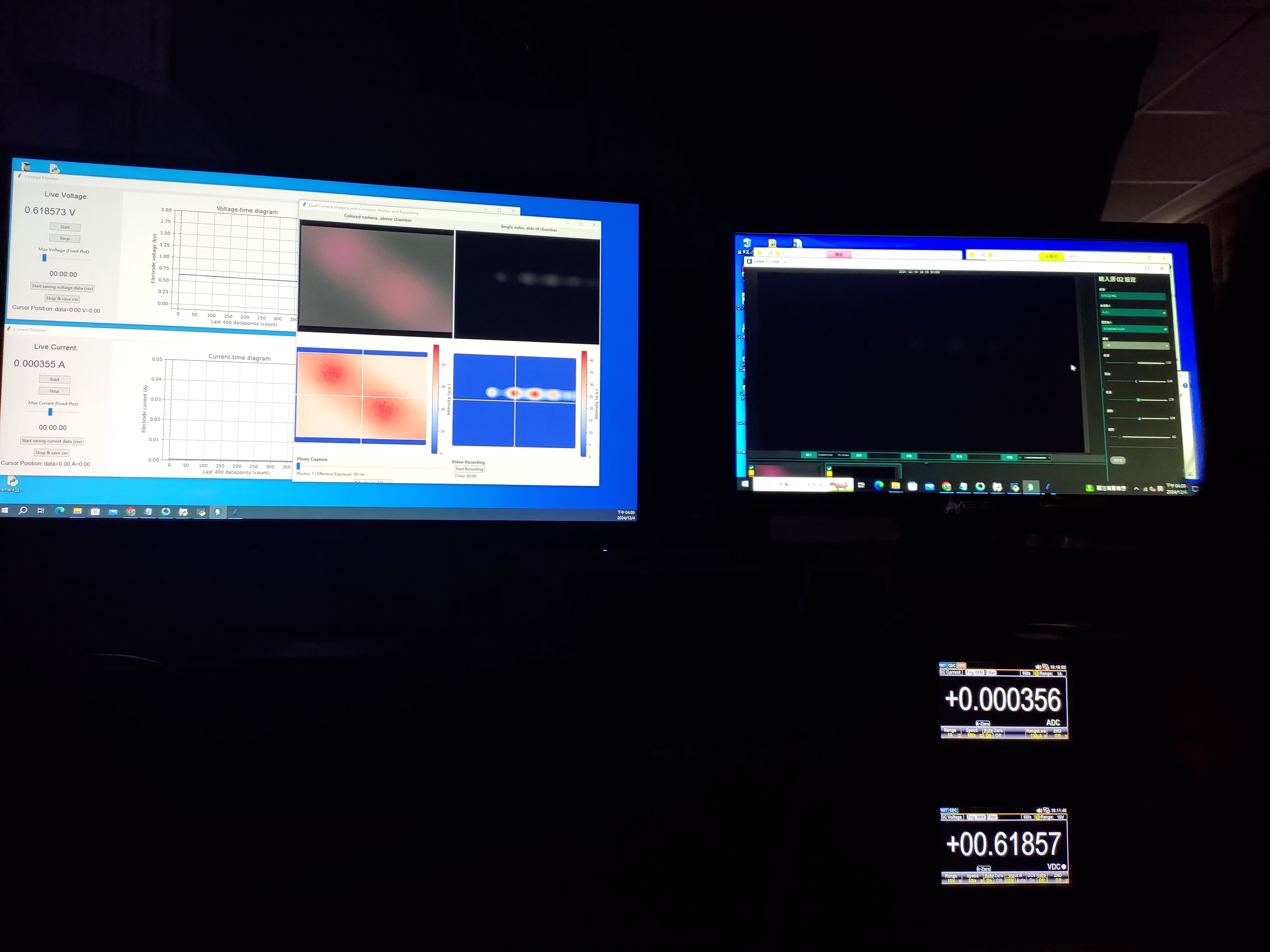}
    \caption{\textbf{Monitoring plasma via GUI during a experiment.} Three GUI's are showed at the left screen. Live voltage and current are displayed on the left. On the right, the two CCDs (colored and gray) are displayed together, with their intensity heat map at the bottom. In comparison, the right is the interface for a commercial software for the CCD, and the live voltage and current shown on the volt-meter and amp-meter.}
    \label{fig:gui}
\end{figure}

With the Python openCV package, we programmed a Graphical User Interface (GUI) for the two cameras and the current and voltage monitors.
For the two CCD camera, the interface includes live streams, live intensity contour map, capture pictures, and capture videos. The exposure time for the CCD cameras are limited to 50 ms, so we stacked many photos together and averaging to achieve exposure of any desired time. As for the current and voltage monitor, we plot a live curve of the measurement with both fixed and scalable ranges. When taking a picture, the camera GUI requests the real-time voltage and current measurement, and output it as the filename for the photo. This design allows for a more efficient data taking process. Fig.~\ref{fig:gui} shows the GUI in operation.

\subsubsection{Homemade parts}
\begin{figure}[H]
    \centering
    \includegraphics[width=\textwidth]{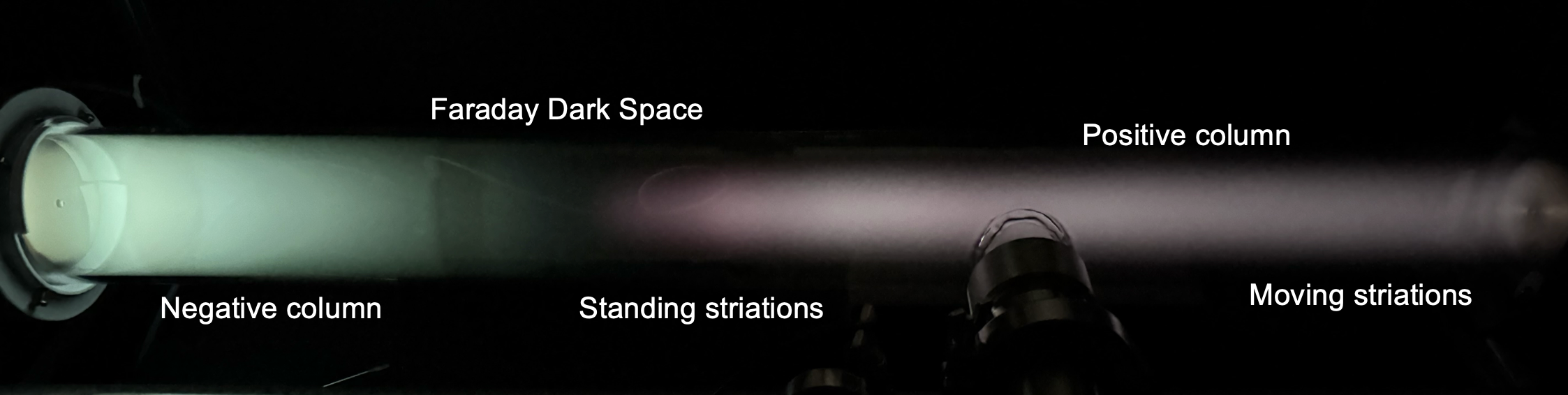}
    \caption{\textbf{Helium DC glow discharge.} \\}
    \label{fig:plasma}
\end{figure}

Lots of parts are homemade in our experiment. A list of homemade parts include the copper electrodes, needle valve, high voltage attenuator, magnetic coil, vacuum clamp for compression fitting, high-voltage connector, and a few home-made wires. We here mention six of them:

\begin{figure}[H]
    \centering
    \begin{subfigure}[t]{0.32\textwidth}
        \centering
        \includegraphics[width=1.3\textwidth, angle=270]{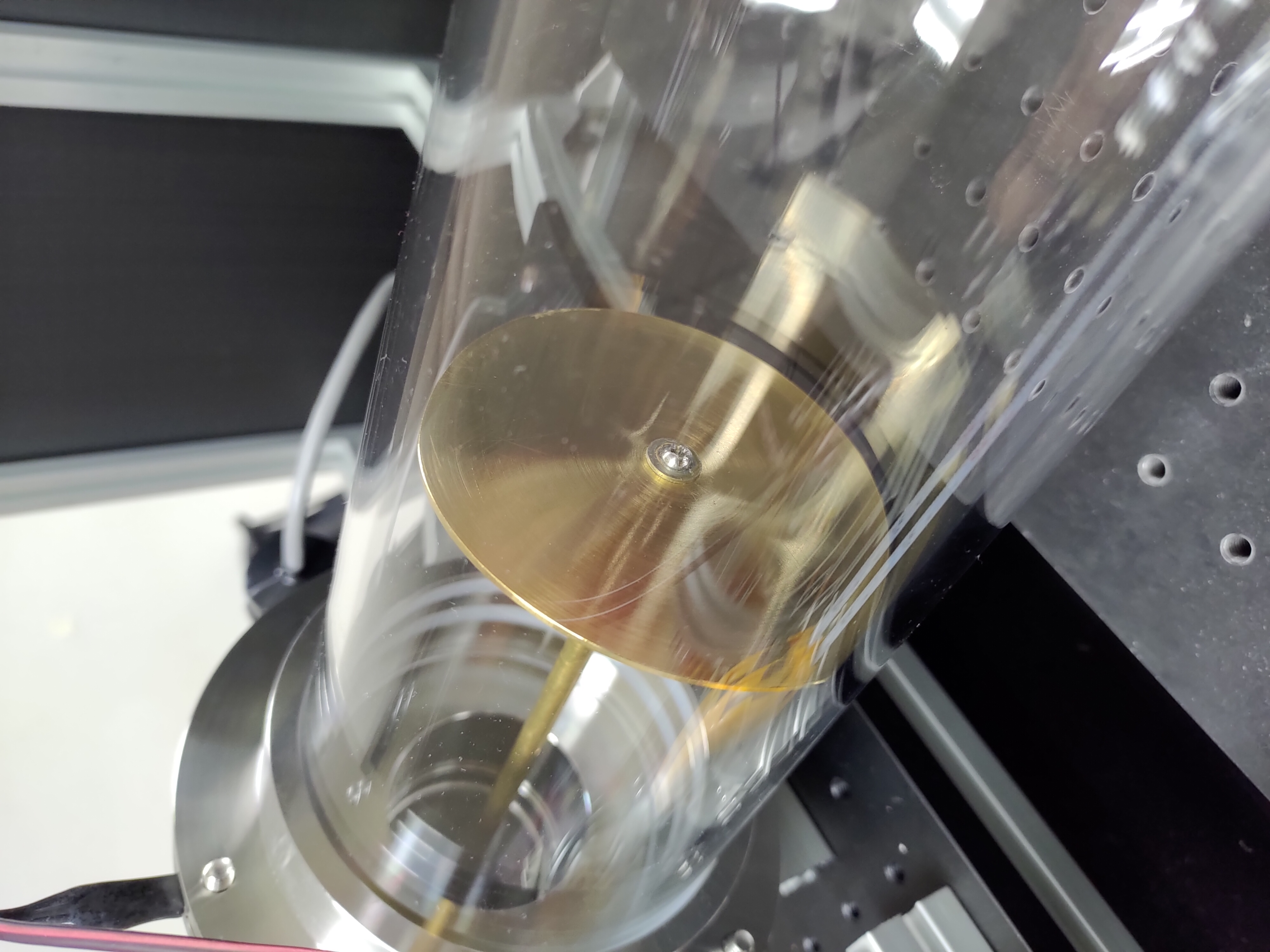}
        \caption{\textbf{Copper electrode.}\\
        A 2mm thick brass disk with a through hole is screwed onto a 1-meter long 1/4" brass rod. Protrusions are designed to ensure electrode plate is perpendicular to rod.}
    \end{subfigure}
    \hfill
    \begin{subfigure}[t]{0.32\textwidth}
        \centering
        \includegraphics[width=1.3\textwidth, angle=270]{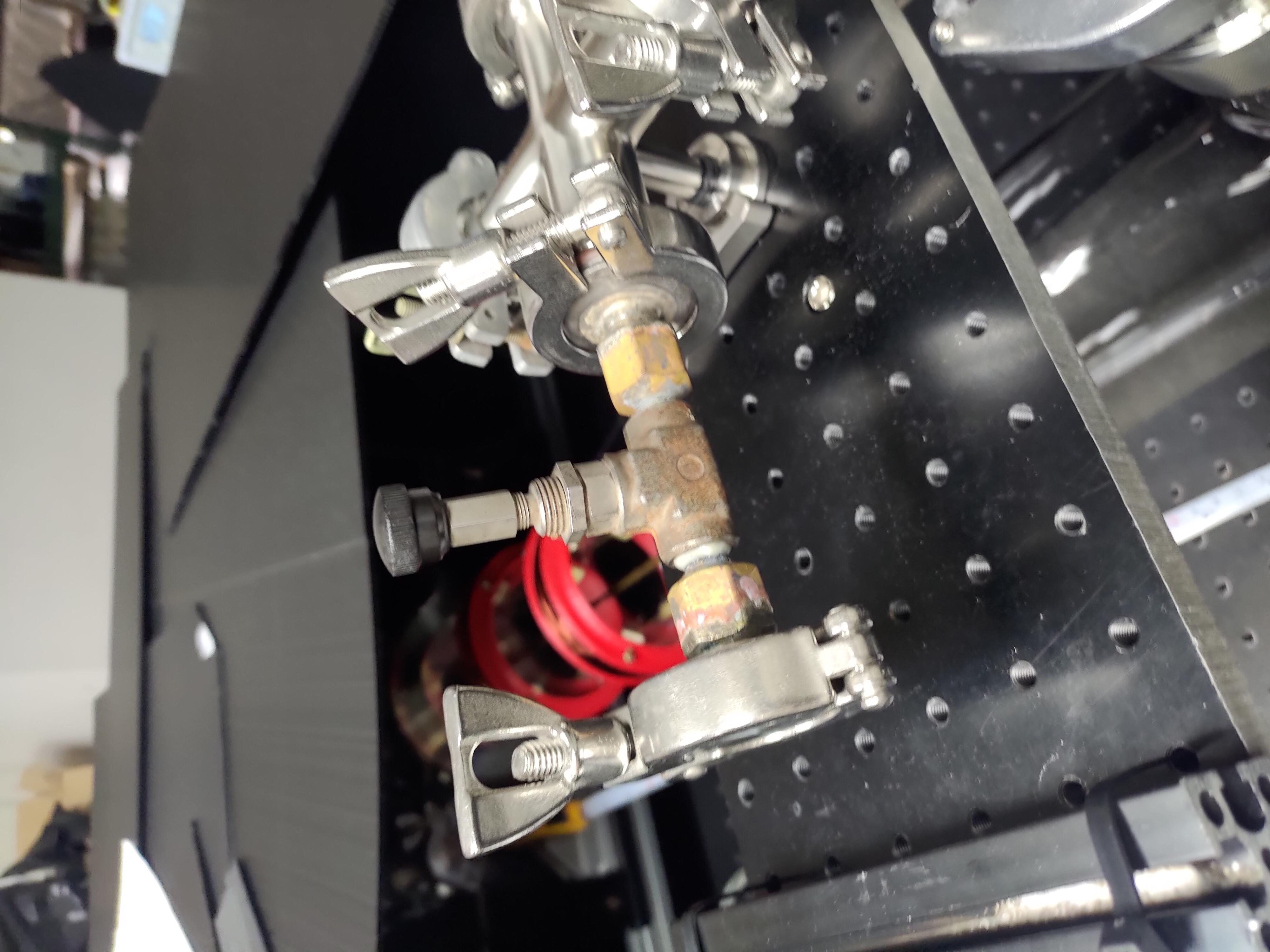}
        \caption{\textbf{Needle valve.}\\
        Two KF25 endcaps are drilled and welded onto a special connector. The connector is then screwed onto the needle valve and sealed by epoxy.}
    \end{subfigure}
    \hfill
    \begin{subfigure}[t]{0.32\textwidth}
        \centering
        \includegraphics[width=1.3\textwidth, angle=270]{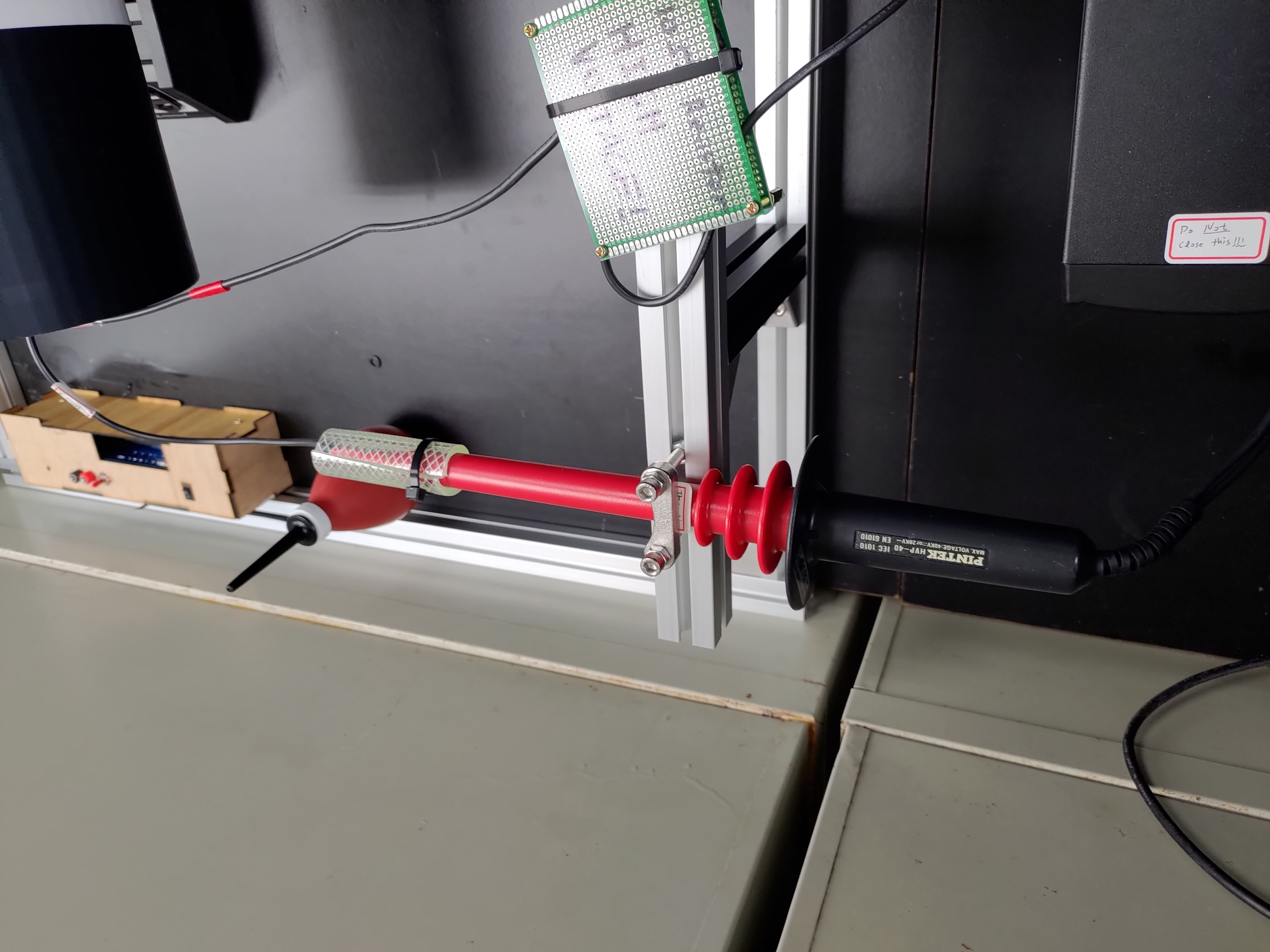}
        \caption{\textbf{High voltage attenuator.}\\
        A 1:1000 high voltage attenuator is rewired to measure the potential difference over the chamber. Everything is covered with plastic or rubber to avoid electric shock.}
    \end{subfigure}

    \begin{subfigure}[t]{0.32\textwidth}
        \centering
        \includegraphics[width=1.3\textwidth, angle=270]{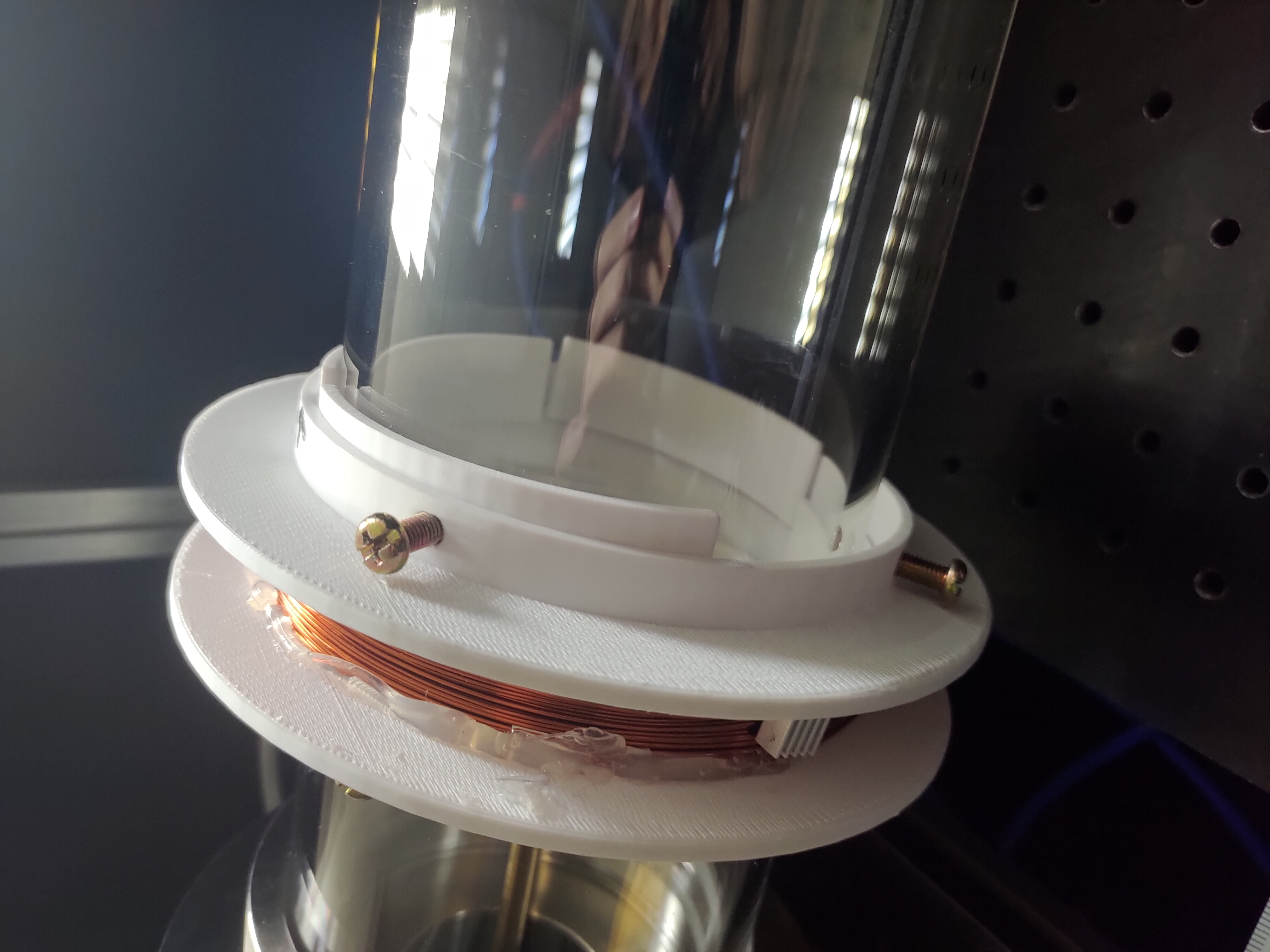}
        \caption{\textbf{Magnetic coil.}\\
        The white structure is printed by 3D printers with PLA material. 100 windings of magnet wire are tightly wound. Further discussion on the magnetic field is in section~\ref{chap:dyn}.}
    \end{subfigure}
    \hfill
    \begin{subfigure}[t]{0.32\textwidth}
        \centering
        \includegraphics[width=1.3\textwidth, angle=270]{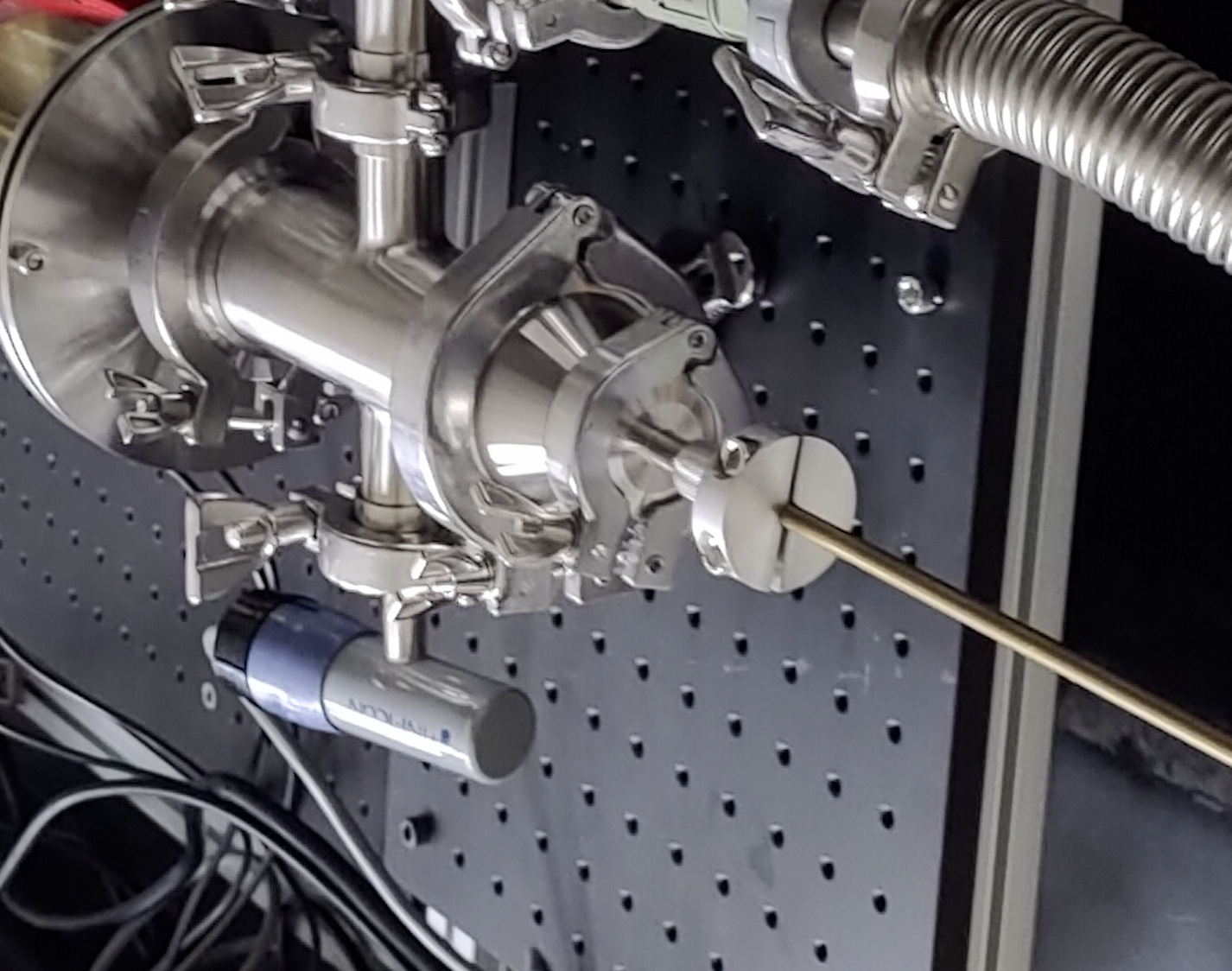}
        \caption{\textbf{Compression fitting clamp.}\\
        The small disk prevents the feedthrough from being sucked into the chamber. A clamp is composed of two aluminum half-rings with inner diameter a bit bigger than the brass rod diameter. The two parts are held together by M6 screws.}
    \end{subfigure}
    \hfill
    \begin{subfigure}[t]{0.32\textwidth}
        \centering
        \includegraphics[width=1.3\textwidth, angle=270]{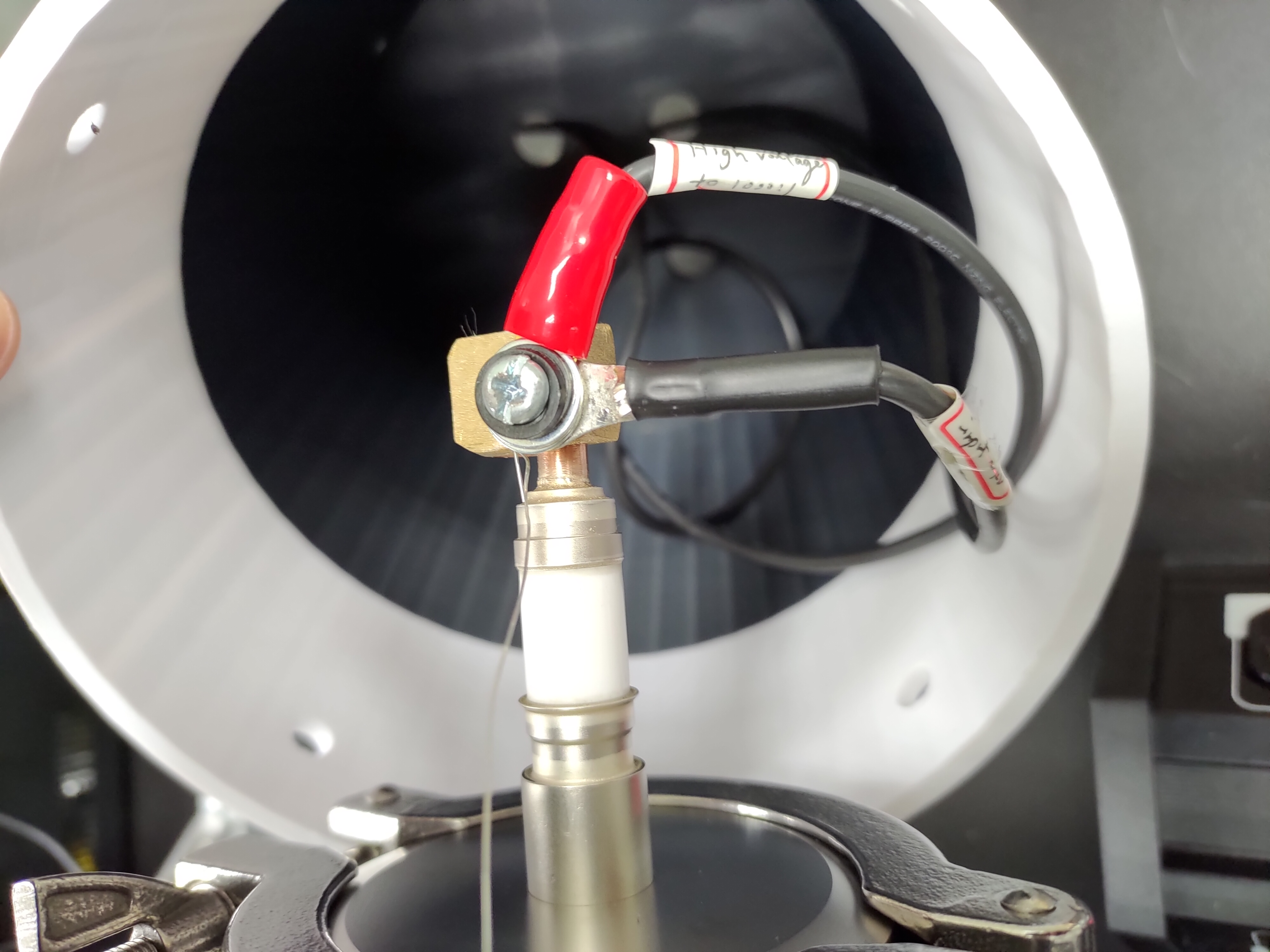}
        \caption{\textbf{Coil connector.}\\
        We sawed a piece of brass with specific matching holes. The wires are screwed onto it and are pressed toward the rod to reduce contact resistance.}
    \end{subfigure}

    \caption{\textbf{A list of homemade parts.}}
    \label{fig:six_images}
\end{figure}
\clearpage
In Fig.\ref{fig:plasma}, a fully constructed plasma chamber with helium dc glow discharge formed at 203 mtorr is shown. We can clearly observe the different structures as previously described.
\subsection{DC Discharge Plasma characteristics}

\subsubsection{High Voltage Breakdown}\label{chap:Paschen}
\begin{figure}[H]
    \centering
    \begin{subfigure}[t]{0.45\textwidth}
        \centering
        \includegraphics[width=\textwidth]{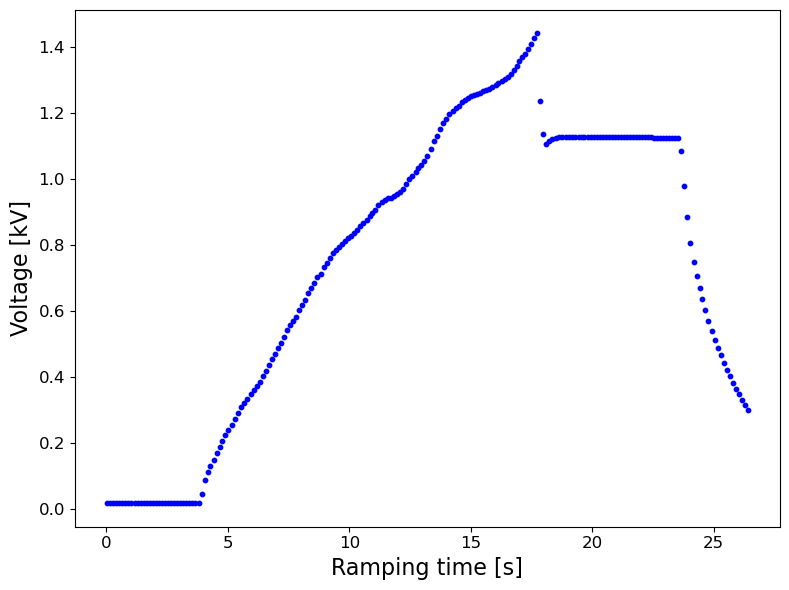}
        \caption{\textbf{The voltage value during the ramp} Plotting the voltage against time, we can see clearly at around 18 s, the gas breakdowns, and the voltage suddenly drops down. The highest voltage value at 18 s is the breakdown voltage }
        \label{'ramp'}
    \end{subfigure}
    \hfill
    \begin{subfigure}[t]{0.45\textwidth}
        \centering
        \includegraphics[width=\textwidth]{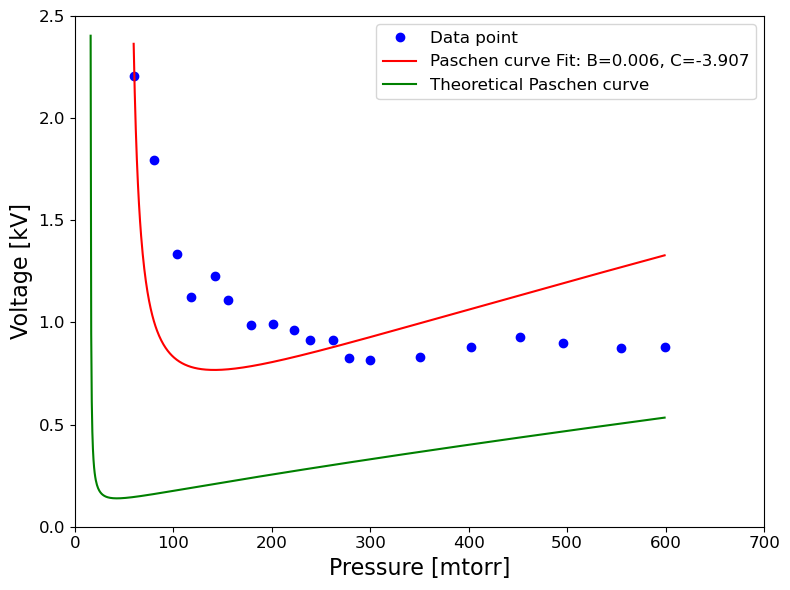}
        \caption{\textbf{The breakdown voltage plotted against pressure} The blue dots are the experimental data. Comparing to the theoretical Paschen curve for helium gas, it requires a high voltage for the gas to breakdown in our experiment. The discrepancy may due to the long cylindrical geometry of the chamber, would need to consider electron diffusion onto the the chamber walls and correct the Paschen curve.}
        \label{'pashen'}
    \end{subfigure}

    \caption{\textbf{The breakdown of a helium glow discharge}}
    \label{Breakdown}
\end{figure}

Slowly ramping up the voltage from zero, we can measure the breakdown voltage of helium gas. As shown in Fig.~\ref{'ramp'}, we can see a clear breakdown when the voltage drops down. This is when the discharge starts to glow stably. The highest voltage value right before the breakdown is the breakdown voltage.\\
The theory of gas breakdown is the Paschen curve, which takes the form:

\begin{equation}
    V_{Breakdown} = \frac{B(pd)}{C + \ln(pd)}
\end{equation}
\begin{equation}
 C=\ln(A) - \ln\left[\ln\left(1 + \frac{1}{\gamma_{se}}\right)\right]
 \end{equation}
Where p is the pressure, d is the distance between electrode (0.954 meters in our case), $\gamma_{se}$ is second emission rate. A and B relates to the first Townsend ionization rate $\alpha$.
\begin{equation}
    \alpha = Ap e^{-Bp / E}
\end{equation}
The theoretic curve is drawn with the coefficient values from \cite{plasma_lab}. A=0.28 ($m^{-1}mtorr^{-1}$), B=3.4 ($Vm^{-1}mtorr^{-1}$), $\gamma_{se}$ = 0.015, C=-2.71.
The fitted value for our experimental data is B=0.006 $Vm^{-1}mtorr^{-1}$, C=-3.907. There is an obvious discrepancy between our measurement and the results from other groups. We suspect this difference is due to the geometry of the chamber. As mentioned in \cite{Lisovskiy2011}, Paschen curve fits best for cylindrical discharge chamber with $\frac{L}{R} \leq 1$, where $L$ is the length of the cylindrical chamber, and $R$ being the radius of the chamber. In our case, $L=1.1$ meters and $R=4.5$ cm, this has a ratio $\frac{L}{R}=24.4$, way beyond the valid range of the Paschen curve. This discrepancy is due to electron diffusion onto the chamber walls. The Paschen theory of breakdown considers all electrons with enough energy would successfully collide into neutral gas, producing additional electrons. Including the electron loss mechanism, the breakdown voltage would need to be higher than what Paschen curve predicts, which is what we observe in the experiment. More details on the corrected Paschen curve can be seen in \cite{Lisovskiy2011}.

\subsubsection{Plasma Voltage-Current Relation}
\begin{figure}[h!]
    \centering
    \begin{subfigure}[t]{0.45\textwidth}
        \centering
        \includegraphics[width=1.3\textwidth]{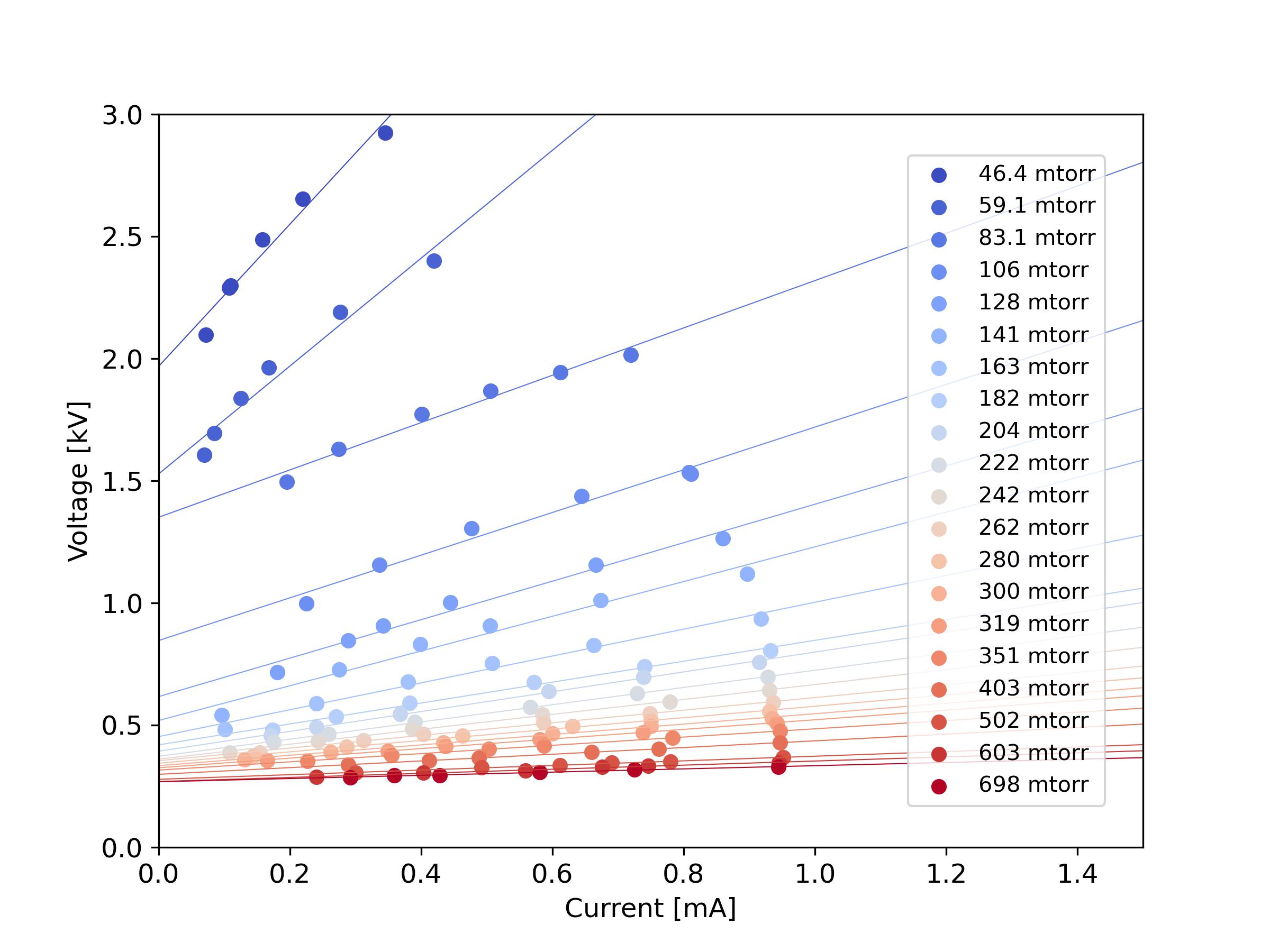}
        \caption{V-I curve of helium discharge at various pressure}
        \label{'vi_curve'}
    \end{subfigure}
    \hfill
    \begin{subfigure}[t]{0.45\textwidth}
        \centering
        \includegraphics[width=1.3\textwidth]{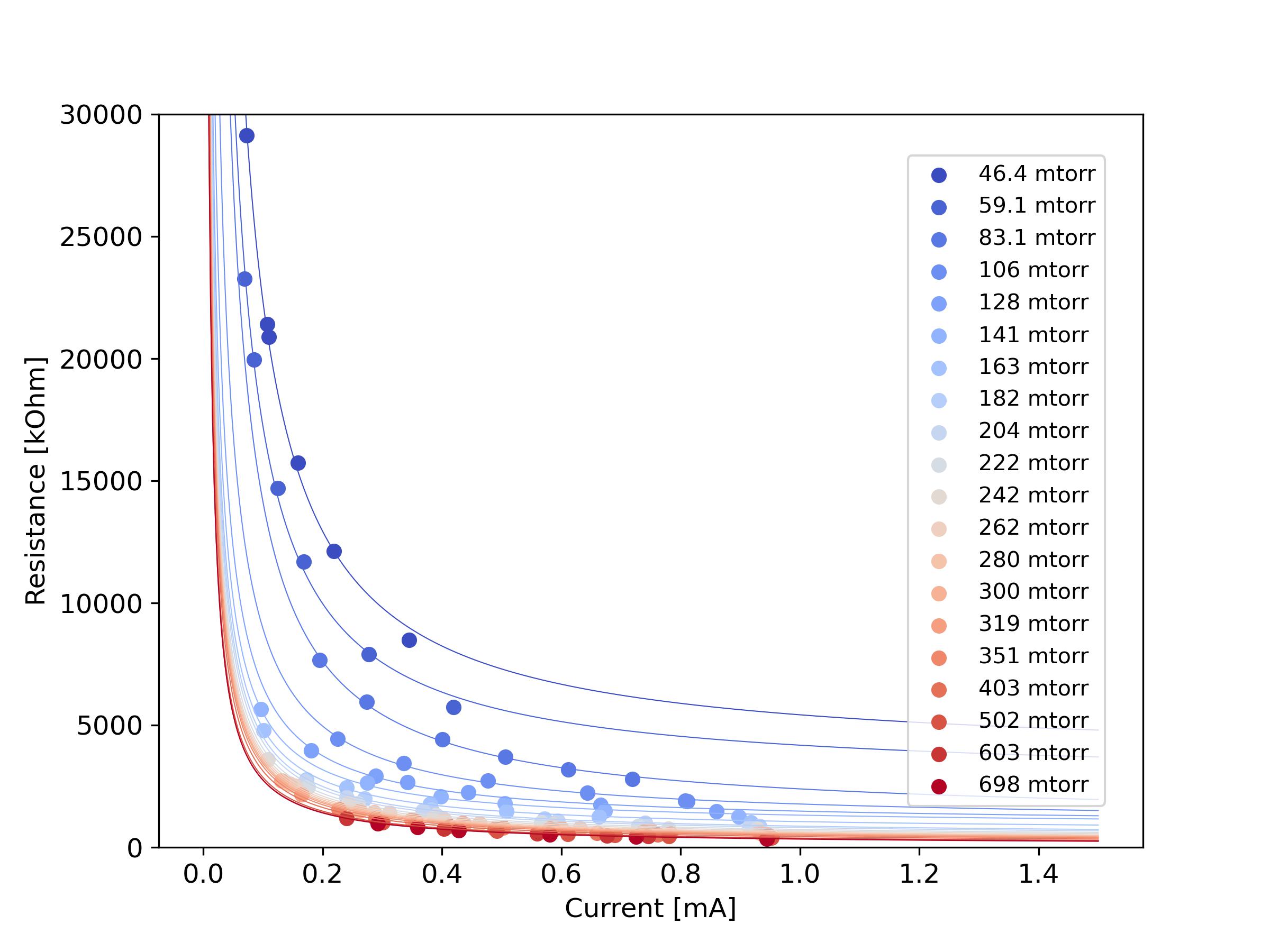}
        \caption{I-R curve of helium discharge at various pressure}
        \label{'ir_curve'}
    \end{subfigure}

    \caption{\textbf{Resistance measurements of a helium glow discharge}}
    \label{Resistance}
\end{figure}

In this section, we study the voltage-current relation of helium plasma under various pressures. In Fig.\ref{Resistance}, we show the voltage plotted against current at helium pressure in the range from 46.4 mtorr to 698 mtorr. We can see that the voltage and current have a linear relation at high pressure, resembling an ohmic conductor with a bias voltage, specifically a diode. This is consistent with the theory mentioned in Fig.~\ref{'IV_ref'} as we are working in the normal glow discharge regime. However at low pressure, the relation slightly differs from an ohmic conductor with bias voltage. We suspect the bias voltage is due to the breakdown nature of the DC discharge. We have to ramp up the voltage to reach breakdown and form self-sustainable plasma as discussed previously. Another explanation is that the voltage drop between the cathode and negative glow draws a constant voltage, as described in~\cite{fundamentals_of_dc}. We used the Child-Langmuir equation~\cite{gonzalez2017childlangmuir} to estimate this bias voltage, and got the correct order of magnitude. So this could also be the origin of the bias voltage.  

By plotting out the resistance against the current, we can see that the resistance ranges from around 1000 k$\Omega$ at high pressure with high current, to 30,000 k$\Omega$ at low pressure with low current. At higher pressure, more atoms can be ionized, leading to more ions to generate second emission. The electrons emitted contribute to the current flow, thus lowering the resistance. The same result was shown in~\cite{princeton_ref} for Argon gas.

\section{Plasma Diagnostics}\label{chap:diag}
\subsection{Langmuir Probe -- Probing Free Electron Temperature and Density}\label{langmuir}
\subsubsection{Theory of Langmuir probes}
Langmuir probes are commonly used as plasma diagnostic tools for determining the free electron temperature and density. A metal probe is inserted into the plasma and biased with a variable voltage, allowing it to collect ion and electron current. Because electrons have a much smaller mass than ions, they exhibit significantly higher mobility and flux. As a result, when a floated conductor is placed in contact with a bulk plasma, electrons reach and accumulate on the conductor more frequently than ions. This leads to the development of a negative potential on the conductor relative to the surrounding plasma. The electrons will then get repelled while ions get attracted until the electron flux equals to the ion flux. The imbalance of charge near the surface gives rise to a localized region called the plasma sheath, which forms to maintain quasi-neutrality in the bulk plasma. To understand the behavior of a Langmuir probe, we must therefore understand the properties of this plasma sheath.

The theory would work under two major assumptions: the plasma is at thermal equilibrium, and the ions are at zero temperature due to its large mass. The energy distribution of electrons would then follow the Maxwell-Boltzmann distribution. 
\begin{equation}
f(E) = 2 \sqrt{\frac{E}{\pi}} \left( \frac{1}{k_B T} \right)^{3/2} \exp\left( -\frac{E}{k_B T} \right)
\end{equation}
Where $k_B$ is the Boltzmann Constant and $T$ is the temperature of the electrons.
\subsubsection{Electric Potential at Plasma Sheath}
We first investigate the potential variation due to the plasma sheath.

\begin{figure}[H]
    \centering
    \includegraphics[width=0.7\columnwidth]{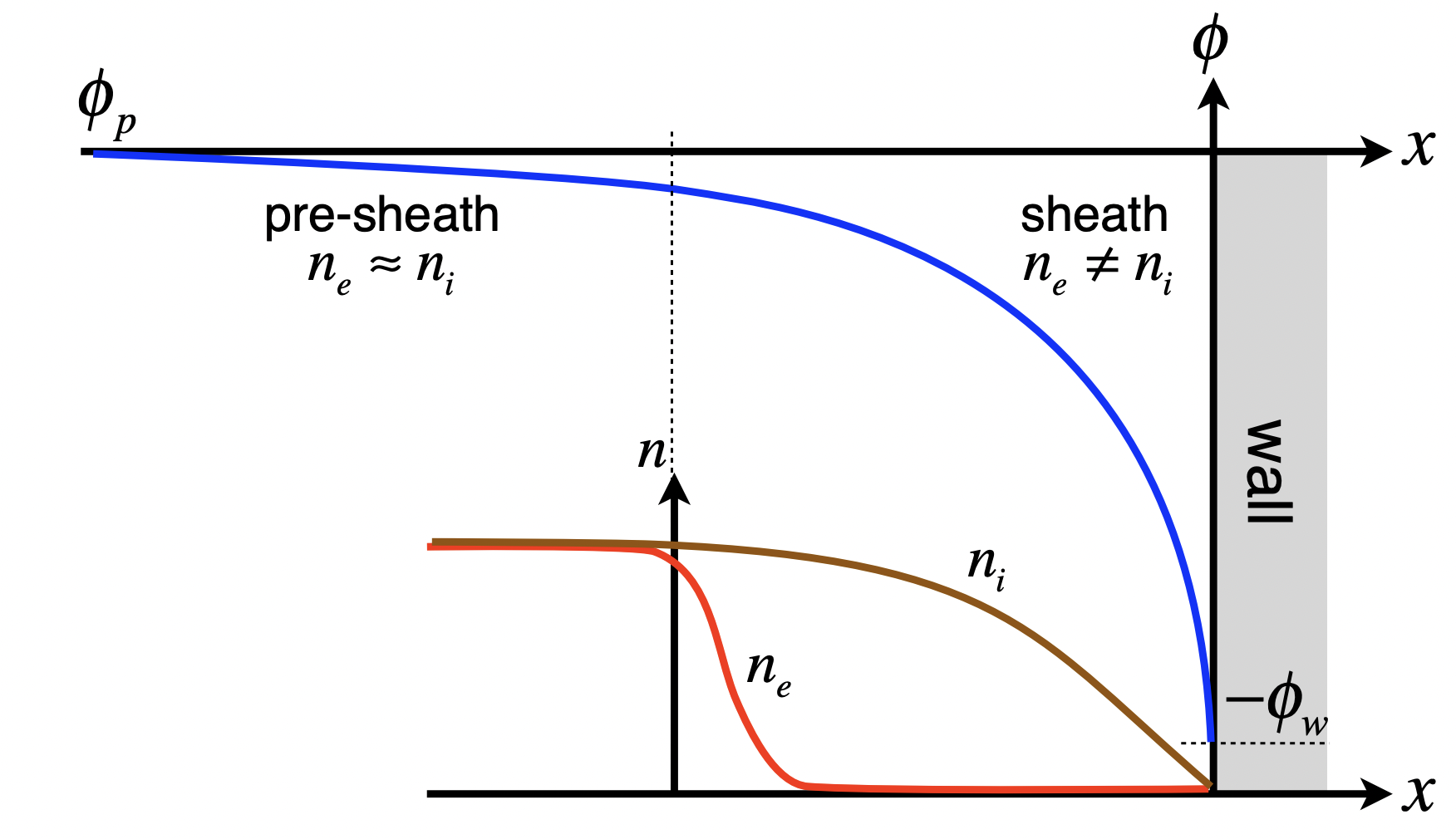}
    \caption{\textbf{Potential and charge variation near plasma sheath.} This graph is taken from \cite{mit-spacepropulsion-lecture9}.}
    \label{'sheath_ref'}
\end{figure}

In Fig.\ref{'sheath_ref'}, the spatial variation of the electric potential from the quasi-neutral plasma to the sheath is illustrated. As shown, the potential drops sharply in the sheath region, becoming increasingly negative toward the conductor surface. This negative potential repels electrons and attracts ions, breaking the quasi-neutrality condition. As a result, the electron and ion densities are no longer equal within the sheath, leading to a region dominated by positive ion charge near the conductor. Electron flux equals ion flux in the sheath, and assuming electron density equals to ion density at sheath edge, we can estimate the potential at the sheath with:

\begin{equation}
V_s = \frac{-k_BT_e}{2e}
\end{equation}
Where $T_e$ is the electron temperature.
\subsubsection{Characteristics of Langmuir Probe Measurement}
\begin{figure}[H]
    \centering
    \begin{subfigure}[b]{0.4\columnwidth}
        \centering
        \includegraphics[width=\linewidth]{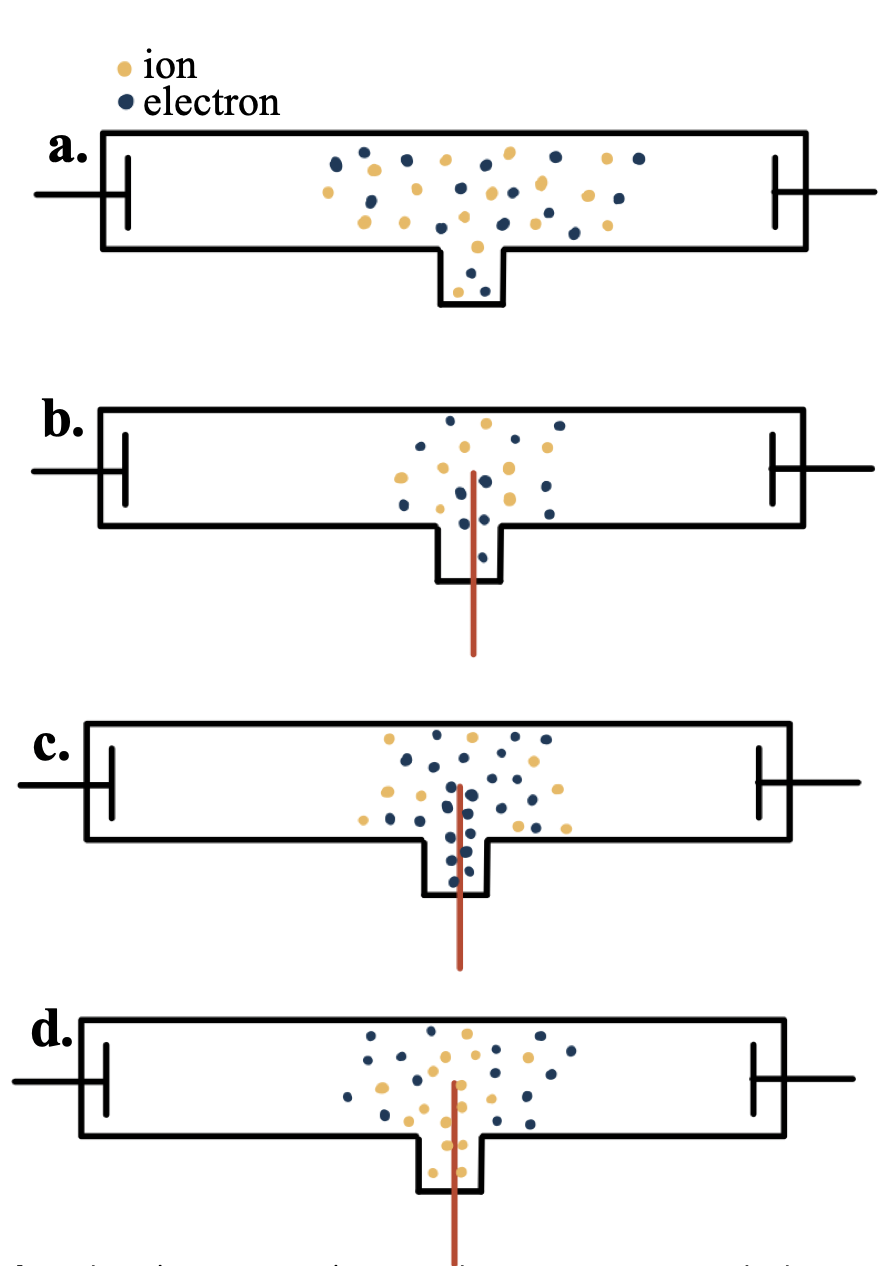}
        \caption{\textbf{Plasma behavior under probing}. a. Plasma consists of neutral gas, electrons, and ions. b. Electrons collide with the probe, creating a negative "floating potential". c. Electron collection regime: a positive bias on the probe attracts energetic electrons. d. Ion collection regime: a negative bias on the probe attracts ions.}
        \label{fig:langmuir_theory}
    \end{subfigure}
    \hfill
    \begin{subfigure}[b]{0.5\columnwidth}
        \centering
        \includegraphics[width=\linewidth]{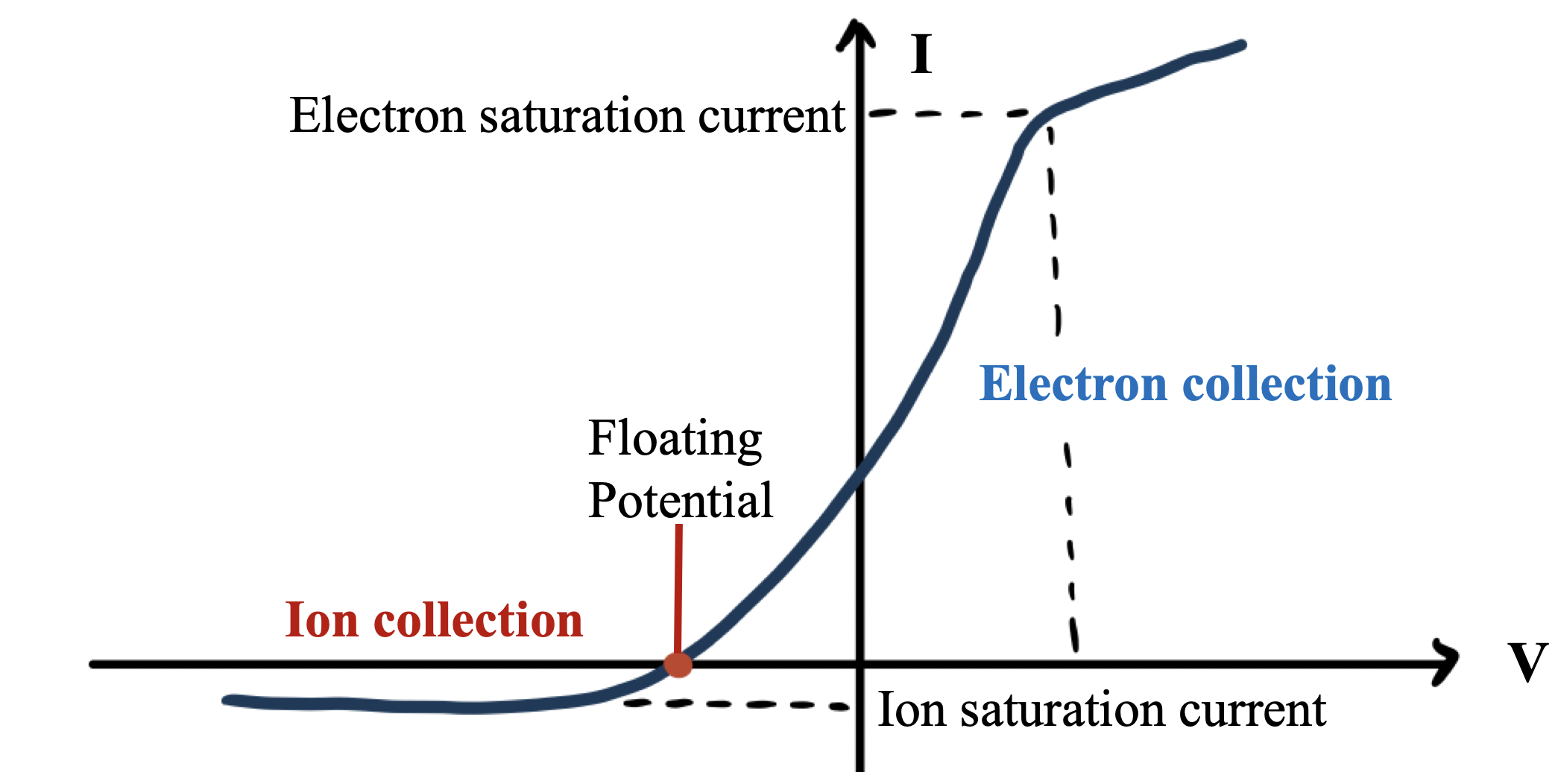}
        \caption{\textbf{Characteristic IV curve}.
         The voltage at which the net current is zero is known as the floating (plasma) potential. For negative bias voltages, the probe attracts ions, resulting in a nearly constant ion saturation current. As the bias voltage becomes positive, the probe begins to attract electrons. In this region, the current increases exponentially with voltage due to the thermal distribution of electron energies, until it reaches the electron saturation current, where nearly all nearby electrons are collected.}
        \label{vi}
    \end{subfigure}
    \caption{\textbf{Theory and characteristic curve of a Langmuir probe.}}
    \label{fig:langmuir_combined}
\end{figure}

During Langmuir probe measurement, we sweep the voltage on the probe, perturbing the plasma sheath. When the voltage is negative relative to plasma potential, the probe repels electrons and attracts ions, draining an ion current. When the voltage is positive relative to plasma potential, it attracts electrons, draining an electron current. By sweeping the bias voltage and measuring the corresponding current, one can obtain the current–voltage (I–V) characteristic of the plasma, from which key parameters like electron temperature and density can be extracted. Fig.\ref{fig:langmuir_theory} illustrates the different current collection regimes as the probe voltage is varied. Tuning the bias voltage and recording the collected current, we can obtain a IV curve that has the characteristic drawn in Fig.\ref{vi}. The I–V characteristic of a Langmuir probe consists of two distinct regions: an exponential region due to electron collection and a constant region due to ion saturation. 

\subsubsection{Derivation of the Langmuir probe IV curve}
Here we derive the form of the collected currents under the sheath theory. The goal is to relate the electron/ion density and temperature at the sheath to the electron/ion in the plasma bulk. At negative biasing, the ion saturation current relates to the probe area A, ion density at conductor $\ n_i(0)$, and the average ion velocity at probe $\bar{u}_i(0)$. 
\begin{equation}
I_{is} = -e A n_i(0) \bar{u}_i(0)
\end{equation}
By applying the continuity equation, the product of ion density and velocity must remain constant between the sheath edge and the conductor surface. Therefore, the ion current can be expressed as:
\begin{equation}
I_{is} = -e A n_i(s) \bar{u}_i(s)
\end{equation}
In the bulk plasma, the quasi-neutrality condition holds, meaning the ion density is approximately equal to the electron density. At the sheath edge, this implies that the ion density equals the electron density at that location. Using the Boltzmann distribution for electron energy, we can relate the ion density at the sheath edge to the sheath potential, electron density and temperature in the bulk plasma.
\begin{equation}
n_i(s) = n_e(s) = n_\infty \exp\left( \frac{e V_s}{k_BT_e} \right)
\end{equation}

We also have energy conservation for the ions traveling through the sheath.
\begin{equation}
\frac{1}{2} m_i \bar{u}_i(s)^2 = e V_s
\end{equation}

\begin{equation}
\bar{u}_i(s) = \sqrt{\frac{k_BT_e}{m_i}}
\end{equation}

With $V_s = \frac{-k_BT_e}{2e}$, we can now derive the ion saturation current in terms of the bulk electron density ($n_\infty$) and temperature ($T_e$).
\begin{equation}
I_{is} =  -0.61\, e A n_\infty \sqrt{\frac{k_BT_e}{m_i}}
\end{equation}
Next we will derive the electron current. When we apply bias voltage V on the probe, only electrons that have energy greater than $e(V-V_p)$ can cross over the potential barrier and reach the probe. We can calculate the density of collected electrons by integrating the Maxwell-Boltzmann distribution.
\begin{equation}
n_e(V) = n_e \int_{e(V - V_p)}^{\infty} f(E)\, dE
\end{equation}
We can then write out the electron current in the same way, including a 0.25 factor due to the isotropic thermal motion of the electrons. Once again using continuity equation we get:
\begin{equation}
I_e = \frac{1}{4} e A n_e(0) \bar{u}_e(0) = \frac{1}{4} e A n_\infty \exp\left( \frac{e V}{k_BT_e} \right) \sqrt{\frac{8 k_BT_e}{\pi m_e}}
\end{equation}

With the two equations of the electron and ion current
\begin{equation}
I_e = \frac{1}{4} e A n \sqrt{\frac{8 k_BT_e}{\pi m_e}} \exp\left( \frac{e V}{k_BT_e} \right)
\end{equation}
\begin{equation}
I_{is} = -0.61\, e A n \sqrt{\frac{k_BT_e}{m_i}}
\end{equation}
We can obtain the electron temperature from extracting the exponent and plug back to the ion saturation current to obtain the electron density.
\subsubsection{Homemade Probe Setup}
We chose molybdenum as the probe material due to its high melting point and high mechanical strength, so it could withstand electron bombardment. We used alumina oxide as a dielectric shield and quarter inch glass tube for structure support. The three layers are sealed together with epoxy to ensure vacuum compatibility. The diameter of the probe is 0.4 mm and the length is 5.52 mm. In addition to the cylindrical probe, we made a planer probe out of copper. The width and height of the probe are 10 mm and 12 mm

\begin{figure}[H]
    \centering
    \begin{subfigure}[b]{0.45\columnwidth}
        \centering
        \includegraphics[width=\linewidth]{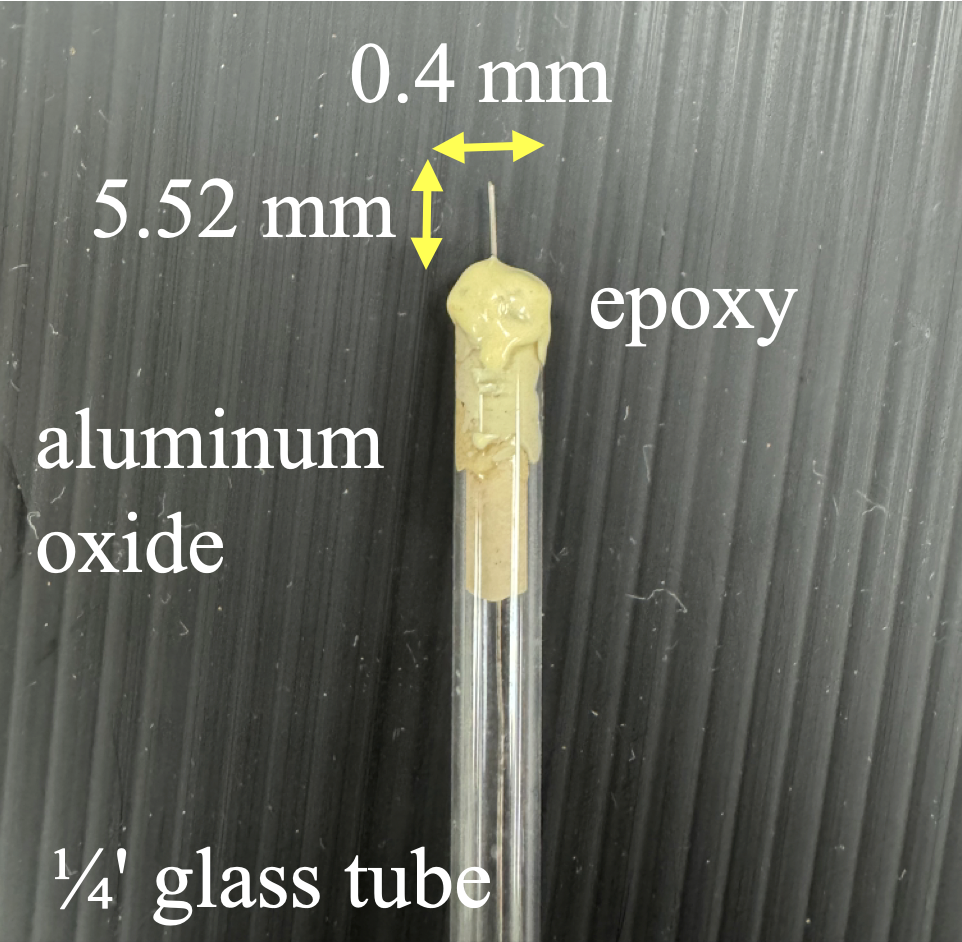}
        \caption{\textbf{Cylindrical probe.}}
        \label{}
    \end{subfigure}
    \hfill
    \begin{subfigure}[b]{0.45\columnwidth}
        \centering
        \includegraphics[width=\linewidth]{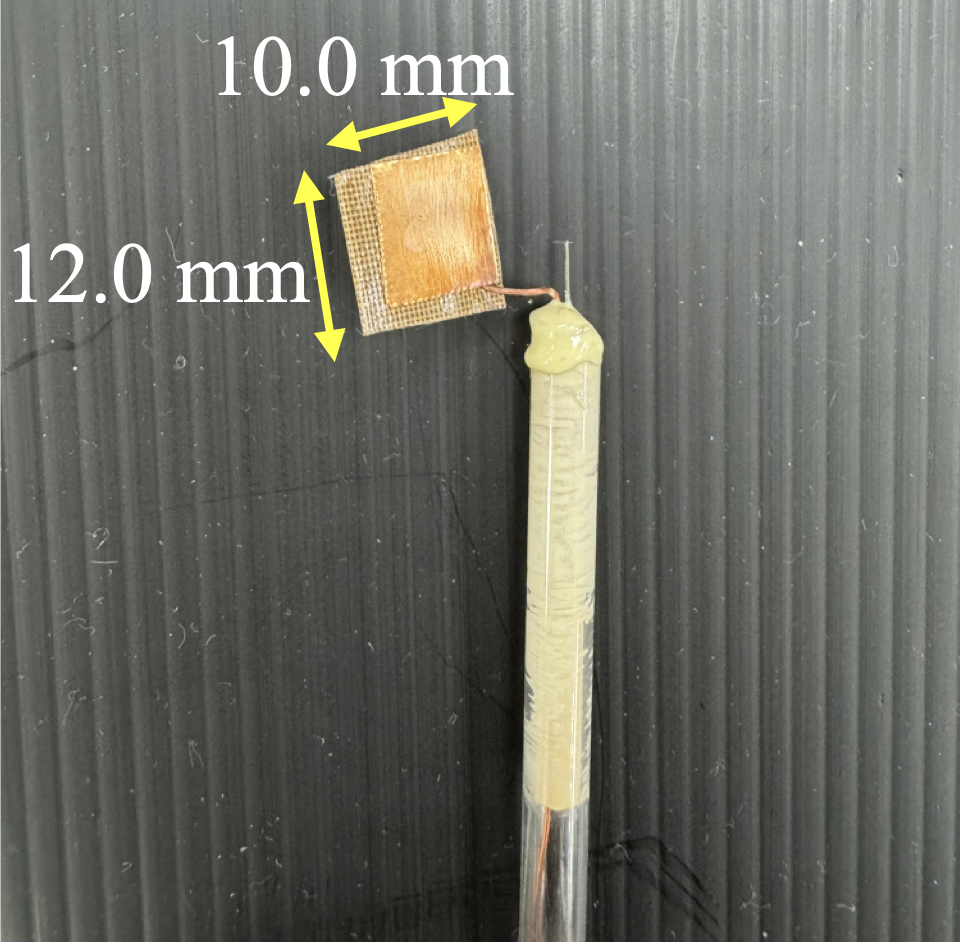}
        \caption{\textbf{Planar probe}.}
        \label{fig:langmuir_vi}
    \end{subfigure}
    \caption{Homemade Langmuir probes. We used aluminum oxide, quartz tube and epoxy for vacuum compatibility.}
    \label{fig:langmuir_probe}
\end{figure}

\subsubsection{Isotropic Test}
\begin{figure}[H]
    \centering
    \begin{subfigure}[b]{0.45\columnwidth}
        \centering
        \includegraphics[width=\linewidth]{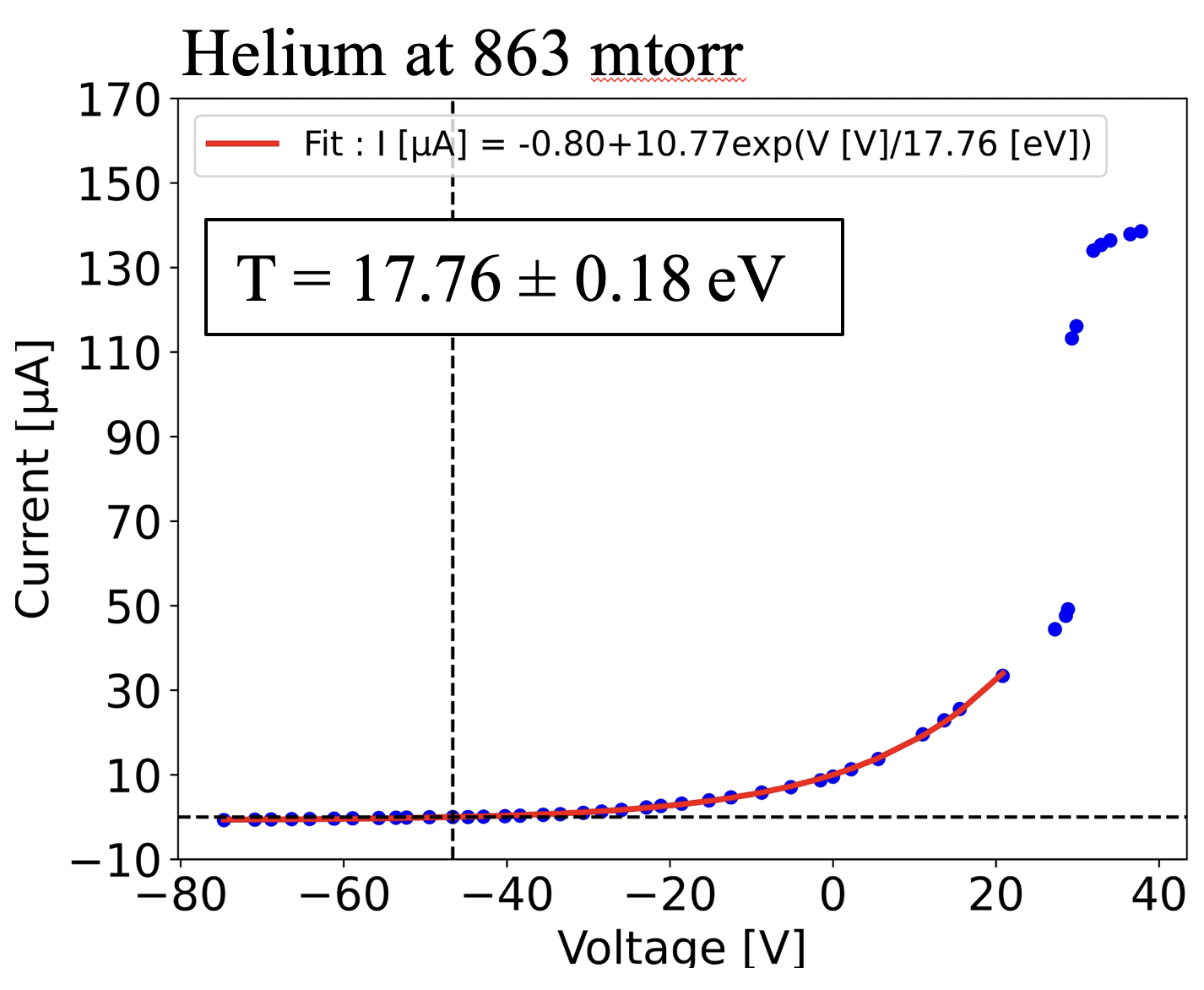}
        \caption{\textbf{IV curve from cylindrical probe.}}
        \label{}
    \end{subfigure}
    \hfill
    \begin{subfigure}[b]{0.45\columnwidth}
        \centering
        \includegraphics[width=\linewidth]{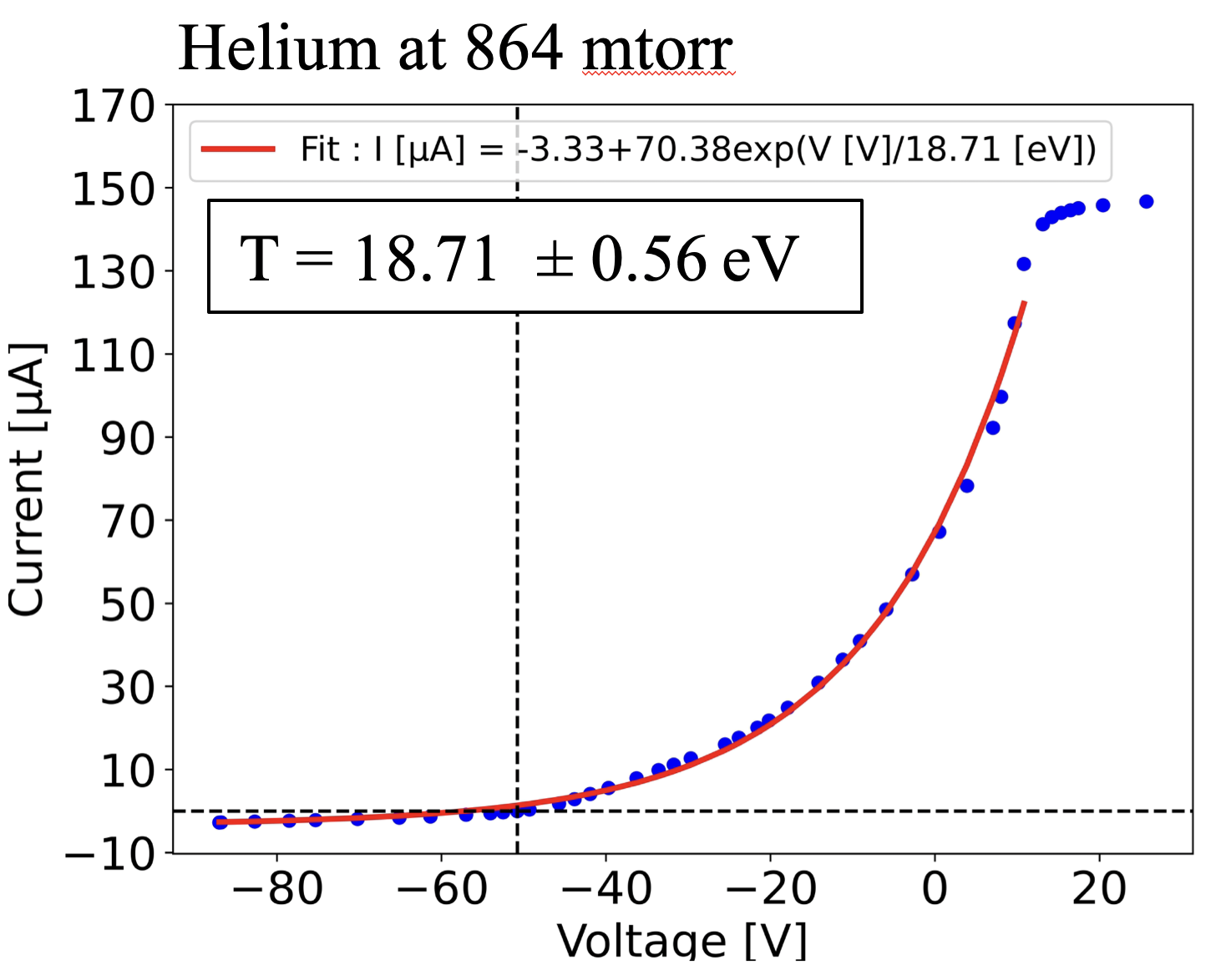}
        \caption{\textbf{IV curve from planar probe}.}
        \label{}
    \end{subfigure}
    \caption{Isotropic Test. We fitted an exponential function with an offset shown as the red line. The two temperature results show good agreement with each other.}
    \label{good}
\end{figure}

Due to the inhomogeneous structure of the DC discharge plasma, and the presence of electron acceleration induced by high-voltage electrodes along the longitudinal axis of the chamber, it is important to first assess whether the plasma deviates from an isotropic Maxwell–Boltzmann distribution. A straightforward approach is to compare measurements from a cylindrical Langmuir probe with those from a planar probe oriented to face the accelerating electrodes directly, thereby probing potential anisotropy in the electron energy distribution.
As shown in Fig.\ref{good}, exponential functions with an offset were fitted to the IV characteristics. From the exponent fitting, we obtained the electron temperature. The cylindrical probe yielded an electron temperature of $17.76 \pm 0.18 eV$, while the planar probe gave $18.71 \pm 0.56 eV$. The close agreement between these values indicates that the plasma remains approximately isotropic, suggesting that the longitudinal acceleration from the electrodes does not significantly bias the electron velocity distribution in that direction.

\subsubsection{Working range of Langmuir probe}
During the measurements, we observed deviations from classical sheath theory when the probe bias voltage became too negative or too positive. Based on these observations, we can establish a operational voltage range for our homemade probes, within the range the extracted plasma parameters remain physically consistent and theoretically valid.
\begin{figure}[H]
    \centering
    \begin{subfigure}[b]{0.5\columnwidth}
        \centering
        \includegraphics[width=\linewidth]{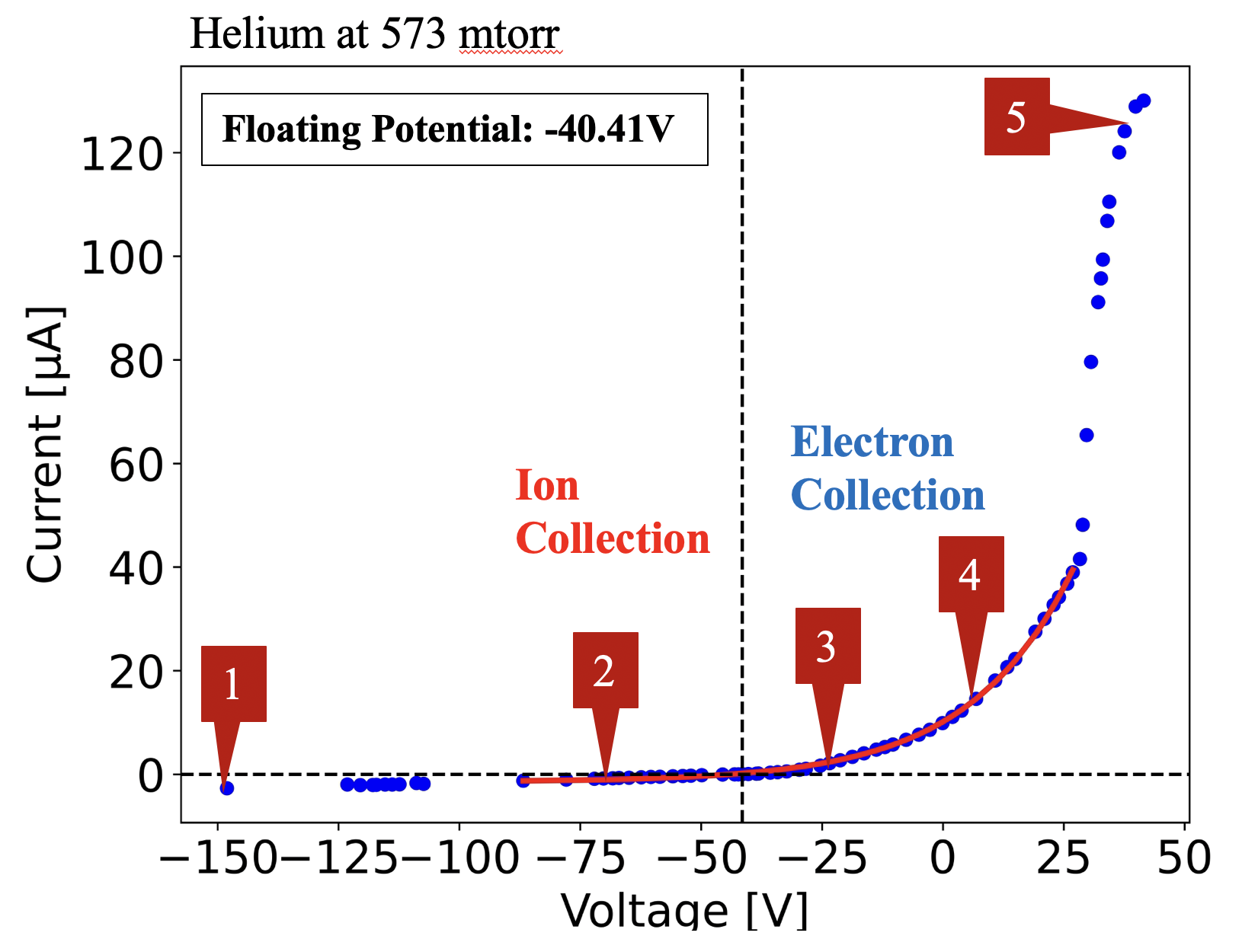}
        \caption{\textbf{IV curve from cylindrical Langmuir probe.}}
        \label{}
    \end{subfigure}
    \hfill
    \begin{subfigure}[b]{0.45\columnwidth}
        \centering
        \includegraphics[width=\linewidth]{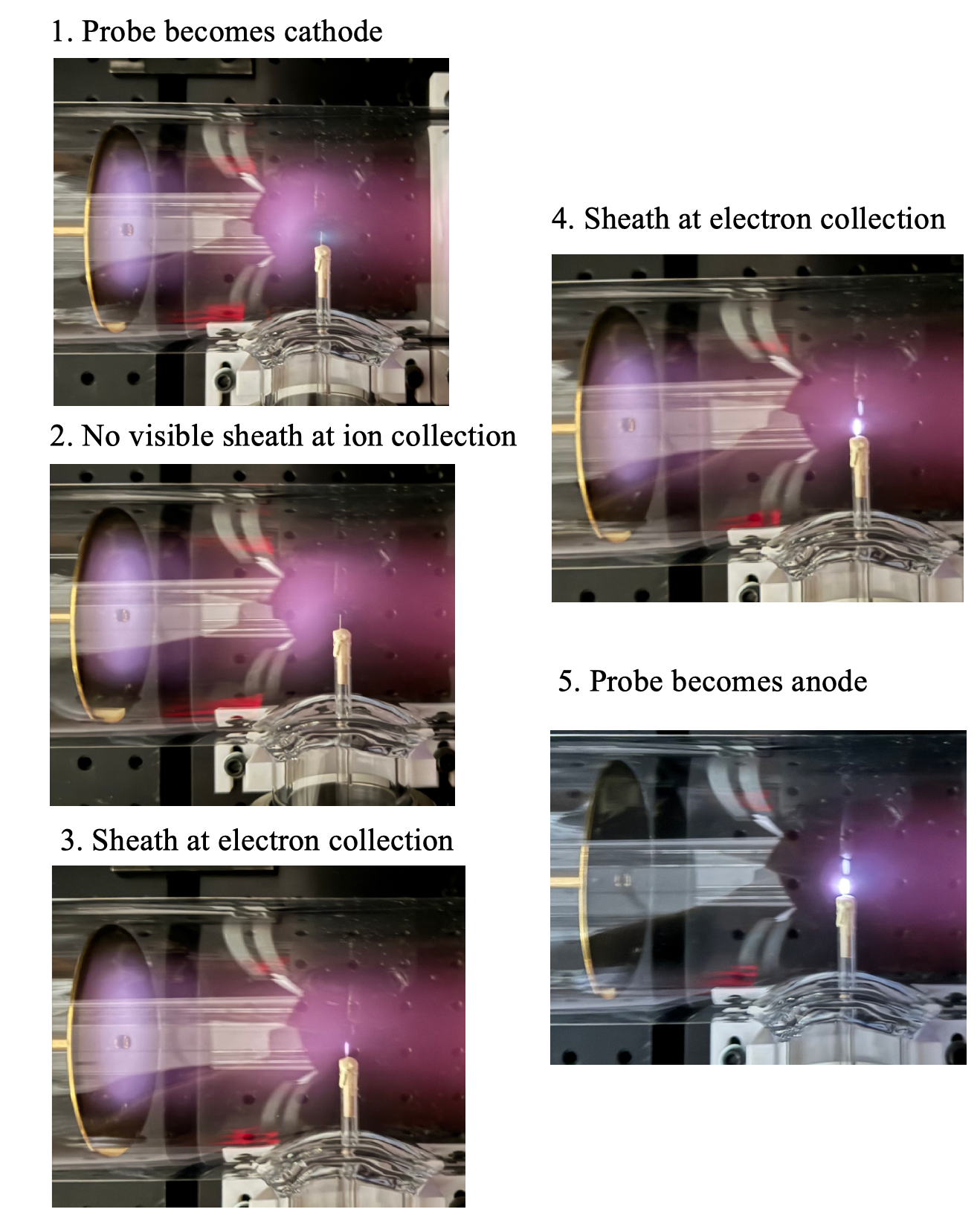}
        \caption{\textbf{Probe behaviors at differnt regions}.}
        \label{}
    \end{subfigure}
    \caption{Working range of Langmuir probe.}
    \label{pic}
\end{figure}
In Fig.\ref{pic}, we show images corresponding to five different points along the IV curve. At Point 1, where the probe bias is set to -150 V, a green glow appears around the probe, indicating that the strongly negative bias causes it to behave as a cathode. This is outside the valid operating range, as cathode-like behavior violates sheath theory assumptions. At Point 2, corresponding to the ion collection region, no visible glow is observed around the probe, suggesting normal ion collection without significant perturbation.

At Points 3 and 4, which are in the electron collection region, a clear glowing sheath is visible around the probe. This glow is likely caused by incoming electrons exciting surrounding helium atoms, resulting in visible light emission. Finally, at Point 5, the probe is biased 80V above plasma potential. Here, the original anode on the right visibly loses its glow, indicating that the probe itself is now acting as the anode, which again falls outside the regime where Langmuir probe theory is valid.

Based on these visual observations, we conclude that the reliable operating range for our homemade probe lies between approximately 
-140 V to 30 V. Within this range, the probe behavior obeys the assumptions of sheath theory, allowing accurate interpretation of the IV characteristics.
\subsubsection{Radial Profile of DC discharge plasma}
\begin{figure}[H]
    \centering
    \begin{subfigure}{0.48\columnwidth}
        \centering
        \includegraphics[width=\linewidth]{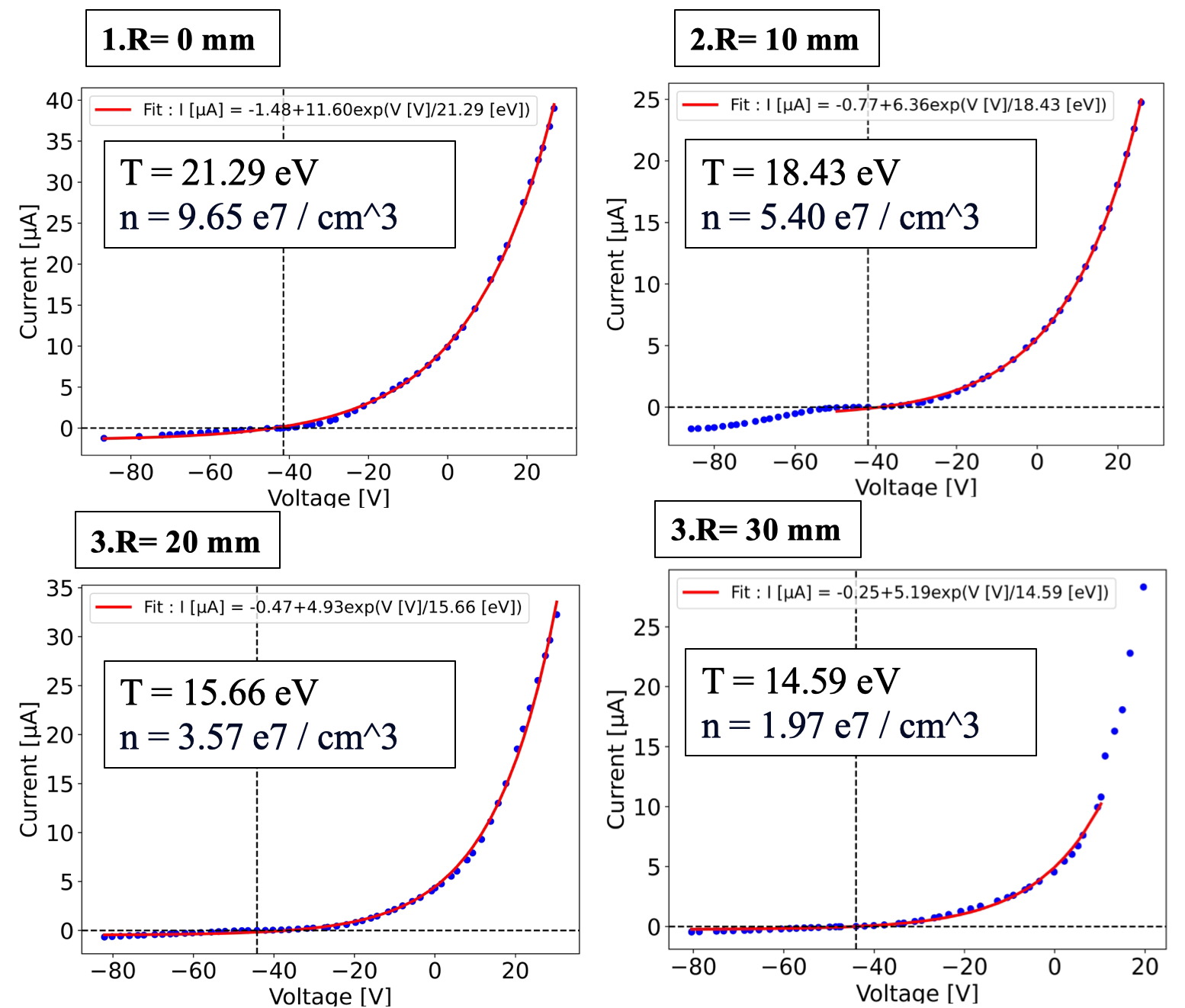}
        \caption{IV curve measurement at different radial positions}
        \label{fig:radialIV}
    \end{subfigure}
    \hfill
    \begin{subfigure}{0.48\columnwidth}
        \centering
        \includegraphics[width=\linewidth]{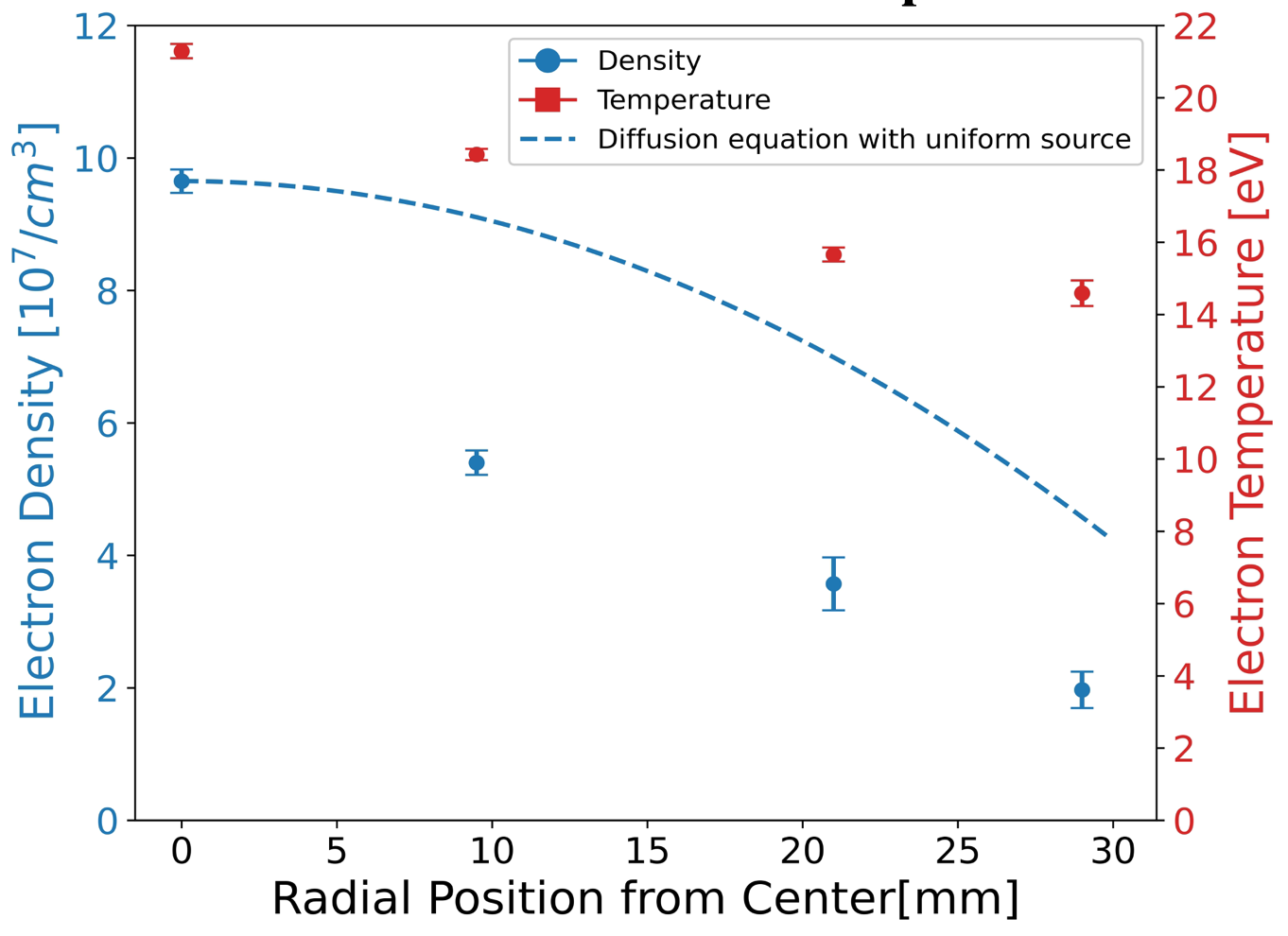}
        \caption{Radial profiles of electron properties}
        \label{fig:radialProfiles}
    \end{subfigure}
    \caption{\textbf{Radial dependence of plasma properties.} (a) IV curve measurements at different radial positions. (b) Corresponding radial profiles of derived electron properties.}
    \label{fig:radialCombined}
\end{figure}

By translating the probe along the radial direction, we obtained spatial profiles of electron temperature and electron density across the plasma column. The cross section of our plasma chamber has a diameter of 80 mm. Measurements were taken at four radial positions: at the center (0 mm) and at distances of 10 mm, 20 mm, and 30 mm from the center. The corresponding IV curves, fitted functions, and the extracted values of electron temperature and density are shown in Fig.\ref{fig:radialIV}.

We plotted the electron temperature and density as functions of radial position. Both quantities show a clear decrease toward the edge of the chamber. When considering only electron diffusion, the resulting radial profile—shown as a dashed line—does not fit the experimental data well. This discrepancy suggests that additional effects, such as ambipolar electric fields and the accumulated charge on the glass cell, must be taken into account to accurately model the plasma behavior\cite{Arslanbekov2021Implicit}.

Lastly, the neutral helium density in our chamber at 573 mtorr is $1.8 \cdot 10^{16}/cm^3$, and the electron density $9.65 \cdot 10^{7}/cm^3$, so we can conclude the ionization rate to be 5.36 ppb.

\subsection{Boltzmann Plot -- Probing Electron Excitation Temperature}
\subsubsection{Theory of Boltzmann Plot}
Another way to probe the electron temperature in plasma is through emission spectroscopy. The be exact, the "electron temperature" here describes the energy the bound electron within an atom or molecule. The glow of a discharge plasma arises from excitation and relaxation of electrons in helium atoms. By measuring the emission intensity ratios of different electronic transitions that decay to a common lower state, we can infer the population distribution among several excited states, and thus determine the temperature of the bound electrons, also known as the excitation temperature.

\begin{figure}[H]
    \centering
    \includegraphics[width=0.4\columnwidth]{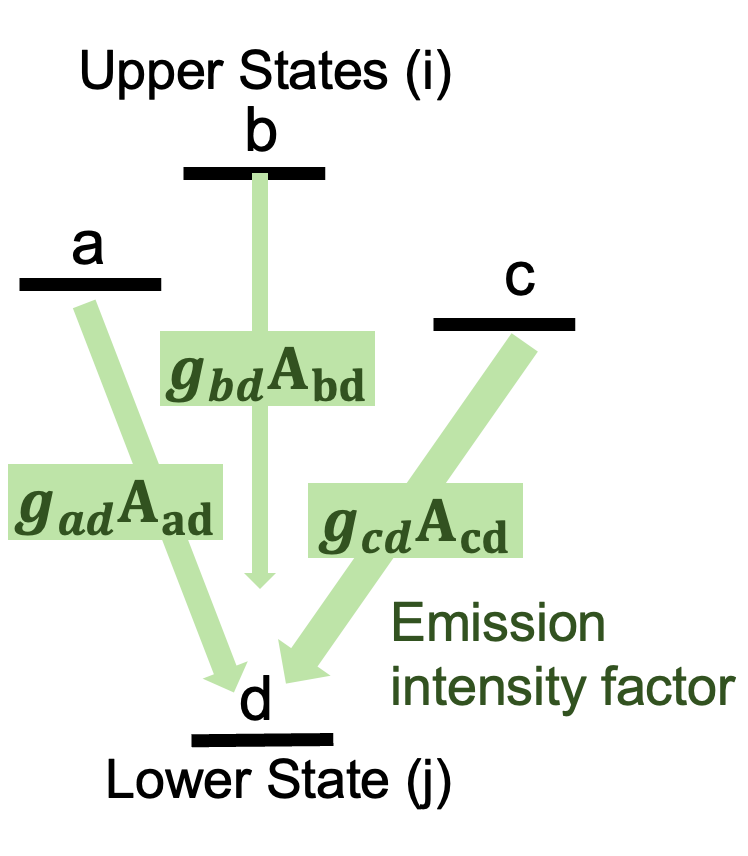}
    \caption{\textbf{Working principle of Boltzmann Plot.}}
    \label{bot}
\end{figure}

Figure \ref{bot} is the working principle of a Boltzmann plot. We chose one common lower state j, and monitor the population across many upper states i by measuring the transition from upper to lower. The intensity is proportional to the occupation fraction $n_i$, $A_{ij}$ Einstein coefficient of spontaneous emission, and $\nu_{ij}$ the frequency of the transition. 
\begin{equation}
    I_{ij} \propto n_i A_{ij} h \nu_{ij}
    \label{int}
\end{equation}
Once again, we assume that the plasma is in thermal equilibrium. Under this condition, the fraction of atoms occupying an electronic state i is given by the Boltzmann distribution:
\begin{equation}
n_i = \frac{g_{ij}}{Z(T)} e^{-\frac{E_i}{k_B T}}
\end{equation}
Where $g_{ij}$ is the degeneracy of state i, $E_i$ is the energy of state i, $Z(T)$ is the partition function, and $T$ is the excitation temperature.

Substituting equation \ref{int} with the Boltzmann distribution gives
\begin{equation}
I_{ij} \propto \frac{g_{ij} A_{ij} h \nu_{ij}}{Z(T)} \exp\left(-\frac{E_i}{k_B T}\right)
\end{equation}
Next, we take logarithm on both sides which will give us the Boltzmann plot relation:
\begin{equation}
\ln\left(\frac{I \lambda}{g_i A_{ij}}\right) = -\frac{1}{k_B T} E_i + \text{const.}
\end{equation}
So if we plot the calculated intensity factor versus the upper state energy, the plot would show a linear relation with slop relating to the excitation temperature.
\subsubsection{Spatial Variation of Excitation Temperature}
\begin{figure}[H]
    \centering
    \includegraphics[width=1\columnwidth]{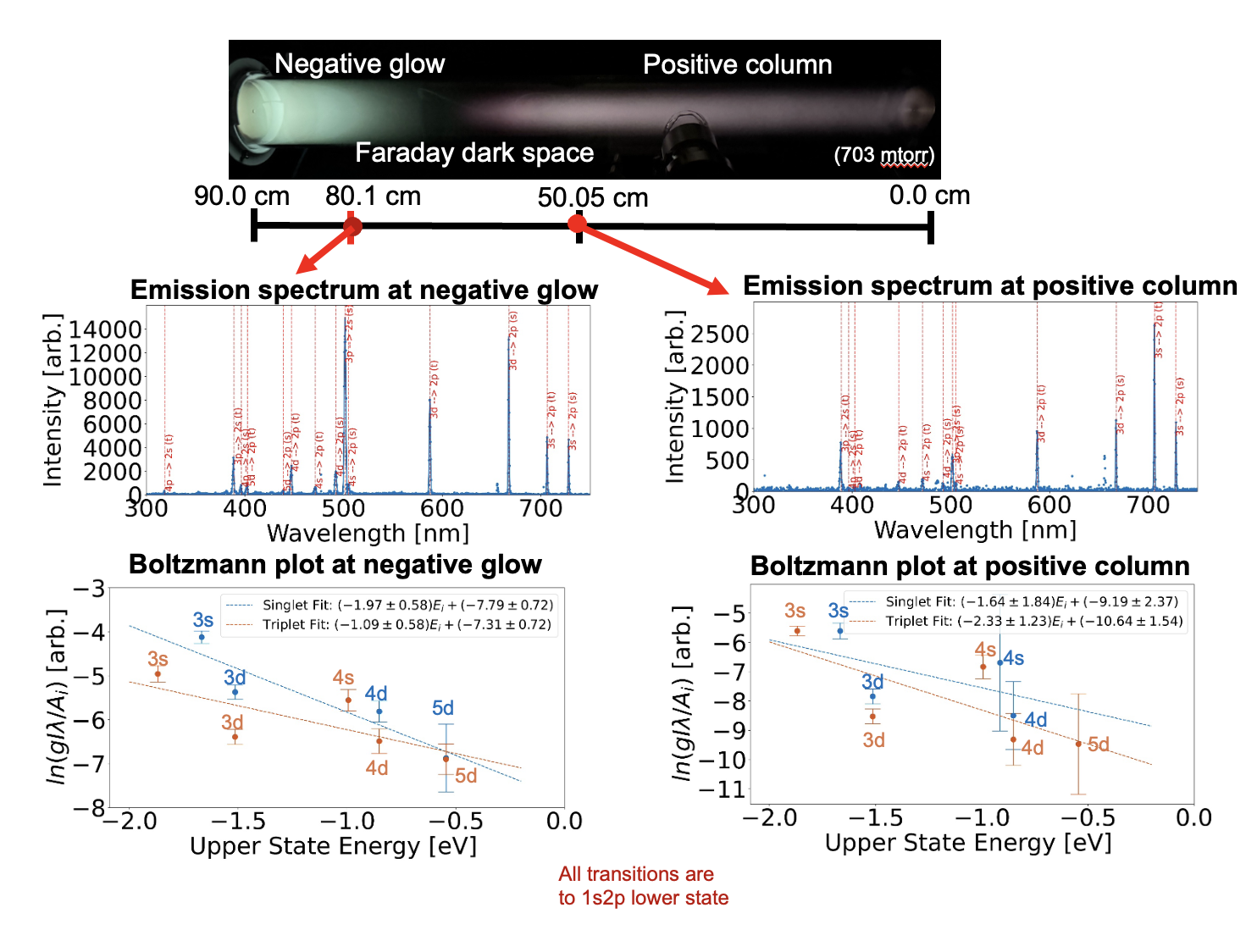}
    \caption{\textbf{Boltzmann Plot of various points in plasma chamber.}The singlet and triplet temperature taken at negative glow is $1.97 \pm 0.58 eV$ and $1.09 \pm 0.58 eV$},respectively. While the temperature taken at positive column is $1.64 \pm 1.84 eV$ and $2.33 \pm 1.23 eV$.
    \label{spectrum}
\end{figure}

We used an Ocean Optics HR4000 spectrometer with a spectral resolution of 0.27 nm to record emission spectra at two regions of the plasma: the negative glow near the cathode and the positive column near the anode. These regions emit distinct colors, corresponding to different spectral features. As shown in Fig.\ref{spectrum}, spectra were collected at 50.05 cm and 80.1 cm from the anode, where the plasma appeared pink and green, respectively. The emission intensity is stronger in the negative glow, dominated by lines near 500 nm, while the positive column shows weaker emission primarily around 700 nm. Each spectral line was fitted using a Voigt profile, and the integrated area under each fitted transition was used to determine the line intensity. We selected transitions from various upper states decaying to the common lower state 1s2p. In the Boltzmann plot, the x-axis represents the upper-state energy, and the y-axis corresponds to the logarithmic term involving the measured intensity. The blue points denote singlet transitions and the orange points denote triplet transitions. The extracted excitation temperatures range from 0.5 eV to 1 eV, with the negative glow region being slightly hotter. The excitation temperature extracted from the Boltzmann plot analysis is found to be lower than the electron temperature measured by the Langmuir probe. Such a discrepancy is expected in a weakly ionized plasma (ionization rate = 5.36 ppm), where electron–neutral collisions are insufficient to establish thermal equilibrium between the electron population and the neutral gas.

\section{Plasma Dynamics}\label{chap:dyn}
\subsection{Magnetic lensing in plasma}
 DC discharge plasma is a weakly ionized gas. As mentioned in section \ref{chap:theory_of_dc}, the electrons accelerate from the cathode to anode, colliding with neutral gas producing ions in the process. If some magnetic field is applied to the plasma, the positively charged ions and negatively charged electrons experience a Lorentz force, which bends their trajectories. For DC discharge, if a magnetic field is present along the trajectory of the electron, it acts like a magnetic lens that focuses the electron beam inwards, reducing the transverse velocity spread. Magnetic confinement is widely used across plasma systems. For example, Tokamaks employ strong magnetic fields to confine plasma and achieve the high densities required for fusion, while magnetron sputtering devices use magnetic fields to trap secondary electrons near the target surface, thus lowering the discharge voltage \cite{Gudmundsson2017}. Overall, magnetic fields are very useful for confining or lowering breakdown voltages in plasma systems.

\subsection{Theory of Magnetic Lens} \label{theory_lens}
Charged particles in a plasma consist of electrons and ions, both of which would be affected by the presence of magnetic field. However, the electron--ion mass ratio is approximately $10^{-4}$, the Lorentz force on the electron is $10^{4}$ stronger than that of the ion with the same velocity. Therefore, the effect of a weak magnetic field of about $50\,\mathrm{G}$ on the ions is negligible. 
In contrast, electrons initially moving with drift velocities on the order of $10^{7}\,\mathrm{cm/s}$ to $10^{8}\,\mathrm{cm/s}$, experience a significant Lorentz force, causing their trajectories to bend.

Magnetic lenses are used in a variety of fields of physics to focus both neutral \cite{Huntington2023} and charged particles. In particular, electron microscope uses a magnetic field aligned to the electron beam's momentum axis to focus the beam to some point determined by the strength of the magnetic field. This is the same configuration that we study here. A commonly used equation that describes the focal length $f$ of a magnetic lens focusing particle with charge $q$ , momentum $p$, with no initial orbital angular momentum with respect to the axis of the lens is \cite{royer1999solenoidal}:

\begin{equation}\label{focus}
\frac{1}{f} = \int \left[ \frac{q B(z)}{2p} \right]^{2} \, dz
\end{equation}

The expression is integrated over the effective range of the lens. In general, the magnetic field forms a harmonic potential inside the lens and if the length of the lens is engineered correctly, all the charged particles can be focused onto the focal point. This technique is used in electron microscopes to obtain a higher resolution \cite{Pany2014}. 

\subsection{Magnetic Lens Setup}
\begin{figure}
    \centering
    \begin{subfigure}[t]{0.5\textwidth}
        \centering
        \includegraphics[width=\textwidth]{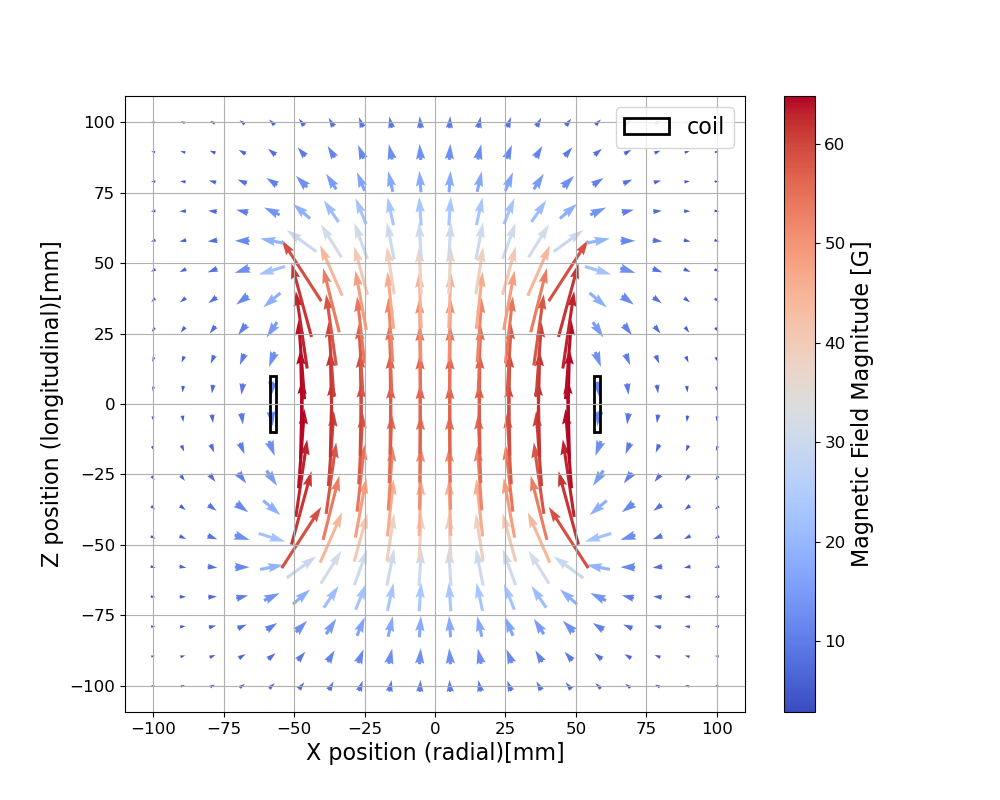}
        \caption{Magnetic field line over a slice of magnetic lens.}
        \label{fig:fieldline}
    \end{subfigure}
    \hfill
    \begin{subfigure}[t]{0.45\textwidth}
        \centering
        \includegraphics[width=\textwidth]{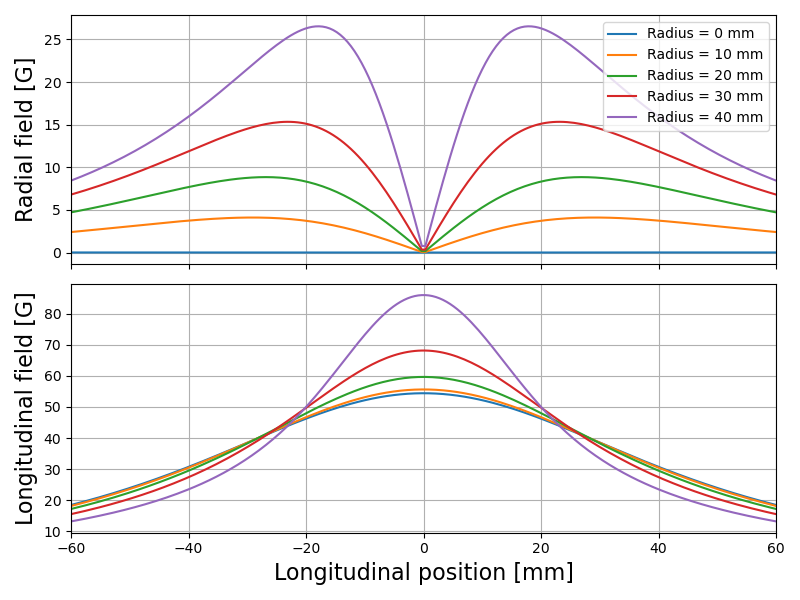}
        \caption{The radial and longitudinal field as a function of z at different radial positions.}
        \label{fig:fieldintensity}
    \end{subfigure}

    \caption{\textbf{Field distribution of the magnetic lens when 5A of current is applied.} The field is minimum at the center of 53 G and increases as a function of radius, reaching 85G at the edge of the chamber. The field for both diagrams are computed numerically by Python.}
    \label{fig:fields}
\end{figure}

We designed magnetic coils to generate a longitudinal magnetic field along the chamber axis. Because the coil’s height-to-diameter ratio is approximately $\tfrac{1}{5}$, its geometry resembles a ring coil rather than a conventional solenoid.
A photograph of the magnetic lens is shown in panel (d) of Fig.~\ref{fig:six_images}, and a similar coil design is used in Princeton Plasma Laboratory's setup~\cite{princeton_ref}. 
In our configuration, the coil produces a magnetic field of up to $50\,\mathrm{G}$ at the center of the chamber, with the field strength increasing radially outward, as illustrated in Fig.~\ref{fig:fields}. 
To confirm the simulated field distribution, we measured the magnetic field using a Gauss meter and found agreement between the measurements and the computed results.\\

\begin{figure}[H]
    \centering
    \begin{minipage}[t]{0.45\textwidth}
        \centering
        \includegraphics[width=\textwidth]{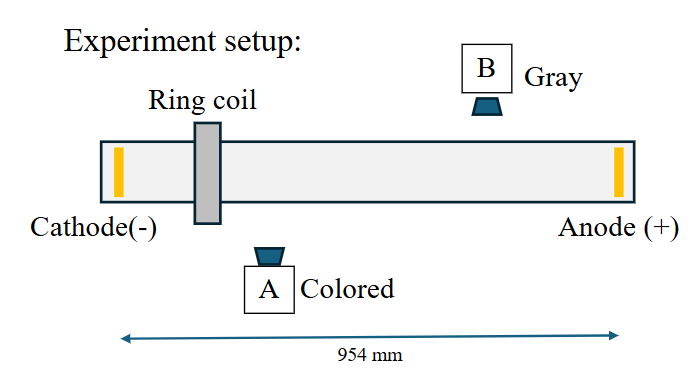}
        \caption{\textbf{Experiment setup for the magnetic lens study.} Two CCD's are setup to monitor the plasma dynamics. The colored one (A) is placed right outside the lens and the gray CCD captures the entire chamber from above.}
        \label{fig:setup}
    \end{minipage}
    \hfill
    \begin{minipage}[t]{0.45\textwidth}
        \centering
        \includegraphics[width=\textwidth]{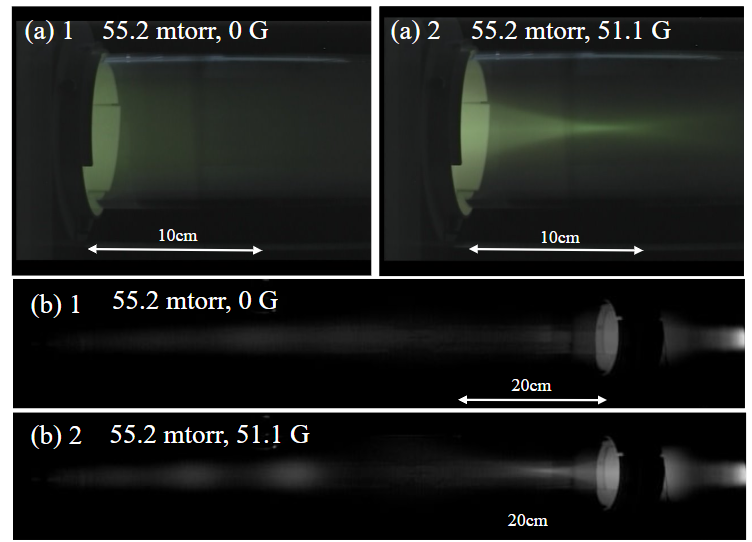}
        \caption{\textbf{Photos captured from the two CCD's with magnetic lens on (54G at center) and off.} The pressure is set to 55.2 mtorrs. (a) Captured by the colored CCD right next to the lens. The green negative glow is tightly focused roughly 14cm away from the lens. (b) shows the focusing significantly changed the electron dynamics such that the plasma property is changed roughly 1 meter downstream near the anode. Initially no striation is visible. After applying the magnetic field, three standing striations are visible. The CCD images are barrel-corrected and calibrated to actual length.}
        \label{fig:typical_lens}
    \end{minipage}
\end{figure}

The experimental setup is shown in Fig.~\ref{fig:setup}. In the experiment, the coil is positioned at various distances from the cathode. When current is applied to the coil, we observe a green light appearing after the lens with a characteristic focusing behavior. This occurs because the electrons are magnetically focused by the coil and subsequently collide with the neutral background gas, producing visible emission. This setup therefore provides a direct visualization of how electron trajectories are bent and focused by a magnetic lens.

\subsection{Measurement Result}

First, we study how the plasma dynamics changes with magnetic field applied at the negative glow. Fig.~\ref{fig:gray_color} presents how the structure changes under different field strengths. The electrons are focused downstream of the magnetic lens, this shows that the particle that is lensed is negatively charged. The only negatively charged particle in the system is the electron. This is what we expect from the theory, as mentioned in section~\ref{theory_lens}.\\

\begin{figure}[H]
    \centering
    \begin{subfigure}[t]{0.5\textwidth}
        \centering
        \includegraphics[width=\textwidth]{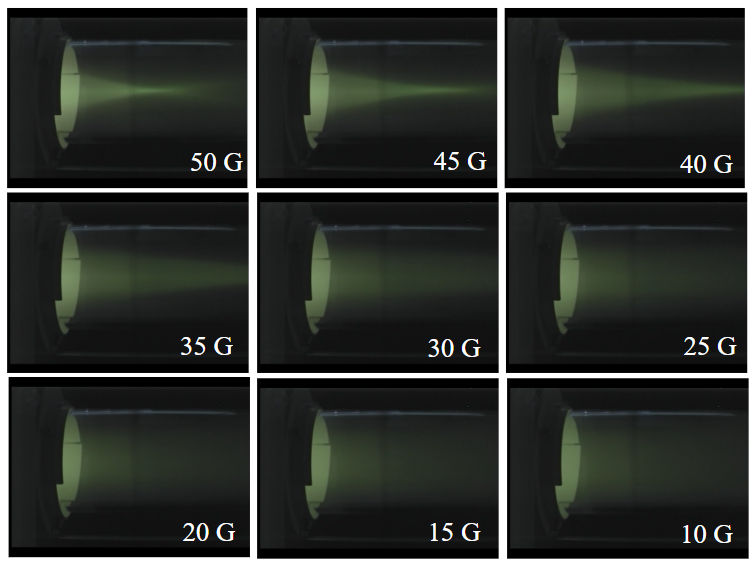}
        \caption{Negative glow right next to the magnetic lens. Captured by colored camera. Cathode is on the left, the black ring is the ring coil. Pictures below 10 G has no observable differences.}
        \label{fig:colored_close}
    \end{subfigure}
    \hfill
    \begin{subfigure}[t]{0.45\textwidth}
        \centering
        \includegraphics[width=\textwidth]{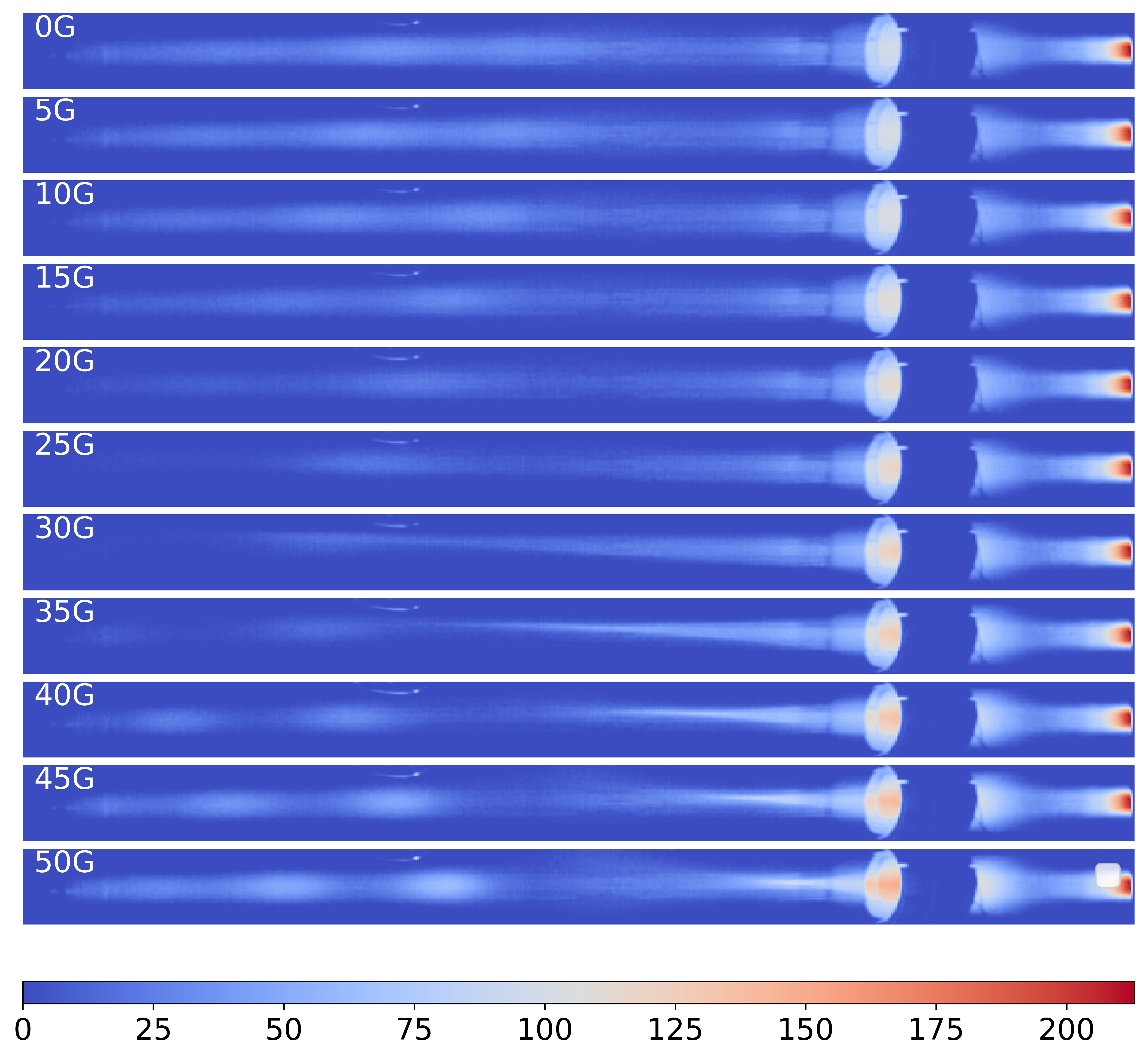}
        \caption{Overall plasma structure under different magnetic field. Cathode is on the right, anode is on the left. The dark area is the magnetic lens. Heatmap colored by intensity.}
        \label{fig:gray_all}
    \end{subfigure}

    \caption{\textbf{Plasma Structures Under Different Magnetic Fields.} From Fig.~\ref{fig:plasma}(a), the intensity at the center of the negative column begins to increase at approximately $25~\mathrm{G}$, and clear beam focusing is observed at $35~\mathrm{G}$. At $45~\mathrm{G}$, the focal point enters the field of view of the colored CCD. From Fig.~\ref{fig:plasma}(b), for $B < 20~\mathrm{G}$ there is no significant change in the negative glow; however, the standing striations shift toward the anode as the magnetic field increases. At $25~\mathrm{G}$, beam focusing becomes visible and the trajectory bends slightly upward, which results from imperfect electrode alignment with the electric field. For $B > 40~\mathrm{G}$, the standing striations reappear and become stronger than those observed for $B < 20~\mathrm{G}$.}
    \label{fig:gray_color}
\end{figure}

We propose a simple theory: The electrons in the negative glow get focused into a narrow beam by the magnetic lens. Since the spatial electron density under lensing has increased, the negative glow gets brighter. This theory can explain all the phenomenon we observe in Fig.~\ref{fig:gray_color}. At weak fields from $0~G < B < 25~G$, the electron beam starts to be successfully focused. Compared to a wider beam, a collimated beam with smaller beam diameter experiences fewer scattering due to the ionization of helium atoms. Now the electron energy is enough to reach the anode, thus the initial striations at the positive column are no longer visible. At $25~G < B < 40~G$, the electron beams are focused such that the focal point roughly land on the anode, so no striations can be observed. At $B > 40~G$, the electrons get focused tightly right next to the lens, and spreads out afterwards. The electrons collide with the electrically-neutral chamber walls and lose energy. In this case, the length of the negative glow region shrinks, and positive column starts to grow. 

This theory can be confirmed by simulations of electron trajectories. Comparing the simulated and measured focal lengths allows us to evaluate whether the theory is correct. The focal length can be obtained from the CCD images by specifying where the electron beam is the most tightly focused. To specify the exact location, we performed a line-cut along the longitudinal direction of the chamber, and search for intensity variations. The method is described in Fig.~\ref{fig:gray_color}. The obtained focal length is listed in table\ref{tab:magnetic_field_data}. In section~\ref{sim}, we will provide the comparison between simulation and experiment.

\begin{figure}[H]
    \centering
    \begin{subfigure}[t]{0.47\textwidth}
        \centering
        \includegraphics[width=\textwidth]{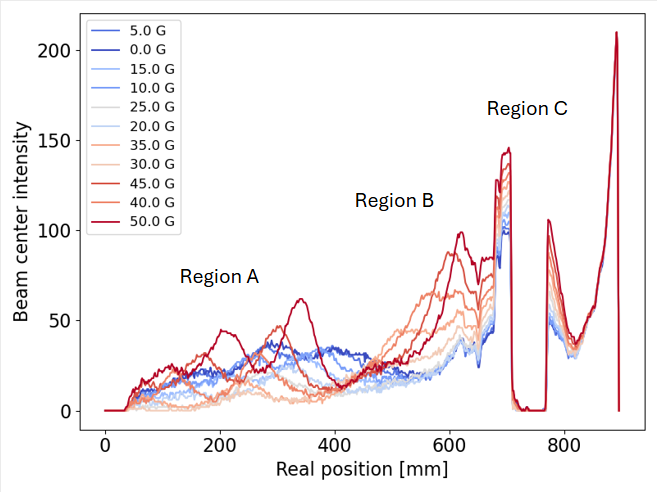}
        \caption{Intensity distribution along the center of chamber of various magnetic field applied at negative glow. }
        \label{fig:intensity_dist}
    \end{subfigure}
    \hfill
    \begin{subfigure}[t]{0.47\textwidth}
        \centering
        \includegraphics[width=\textwidth]{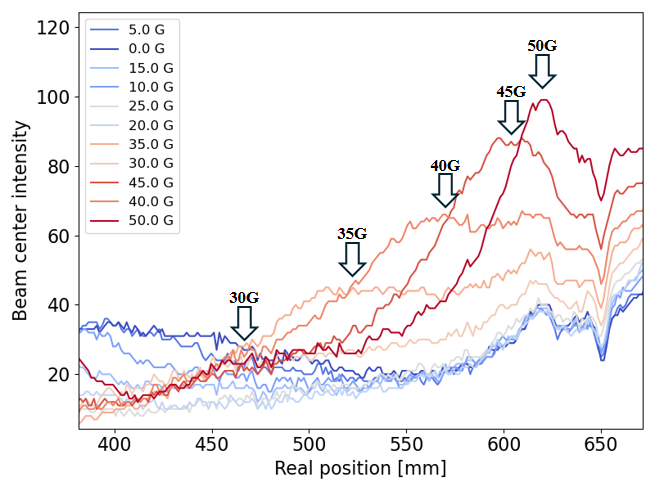}
        \caption{Zoomed into Region B. The arrows specify the focal point of different magnetic field.}
        \label{fig:intsneity_zoom}
    \end{subfigure}

    \caption{\textbf{Intensity distribution under different magnetic fields. The horizontal axis correspond to the distance to the anode.} We can observe the negative glow has a decrease in intensity from cathode to anode as expected. Region A (near anode) is relatively dim at low fields with few striations. When the magnetic field is raised to 30 G, striations disappear. Above that, striations start to reappear with higher intensity. In Region B, the peak between 400 mm to 600 mm correspond to the focal point. In Region C, the light is blocked by the ring coil, resulting in an absorption valley on the diagram.}
    \label{fig:gray_color}
\end{figure}

\begin{table}[H]
    \centering
    \caption{Magnetic field data and associated focal lengths with errors.}
    \resizebox{\textwidth}{!}{%
    \begin{tabular}{ccccc}
        \toprule
        \textbf{Magnetic field (G)} & 
        \textbf{$\pm$ Magnetic field error (G)} & 
        \textbf{Focus (mm)} & 
        \textbf{Focal length (mm)} & 
        \textbf{$\pm$ Error (mm)} \\
        \midrule
        35 & 1 & 533.8 & 245.2 & 30.6 \\
        40 & 1 & 567.7 & 211.3 & 15.4 \\
        45 & 1 & 604.7 & 174.3 & 7.2 \\
        50 & 1 & 618.1 & 160.9 & 4.6 \\
        \bottomrule
    \end{tabular}
    } 
    \label{tab:magnetic_field_data}
\end{table}

\subsection{Magnetic Field Modifications to Current-Resistance Relation.}\label{IR}

\begin{figure}[H]
    \centering
    \includegraphics[width=\textwidth]{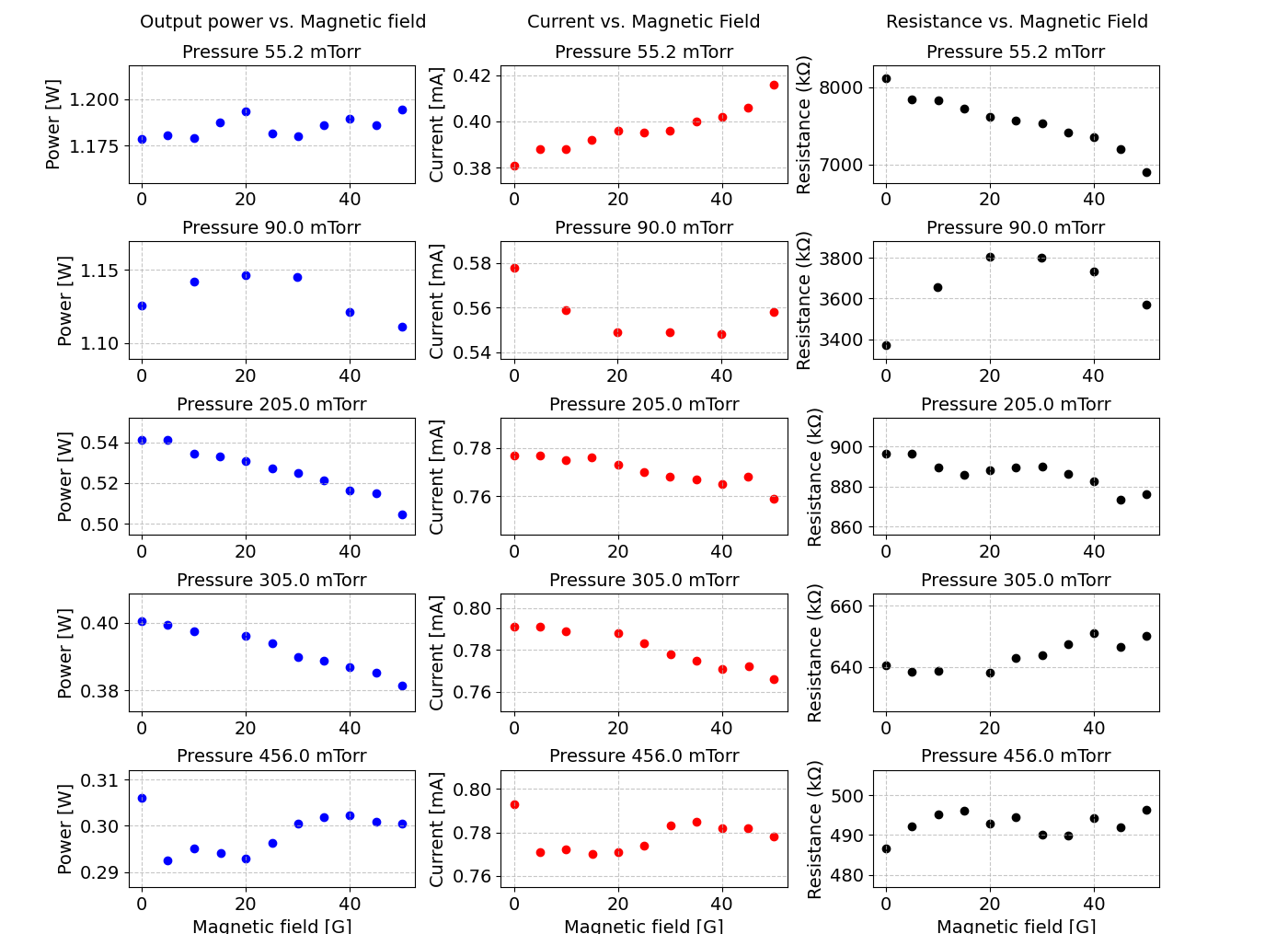}
    \caption{\textbf{Plasma power, current, and resistance under different magnetic fields.} The different rows correspond to different pressures, ranging from 50 mtorr to 450 mtorr.
    During the experiment, the output power of the power supply is unchanged. As can be seen in the "Output power vs. Magnetic field" diagram, the overall change in power is less than 5\%.
    The "Current vs. Magnetic field" and "Resistance vs. Magnetic field" plot show that current and resistance vary with magnetic fields. We see at various pressures, the current-magnetic field can be increasing (55.2 mtorr), decreasing (above 205.0 mtorr), or with a dip (90.0 mtorr). On the other hand, the resistance-magnetic field relation can be decreasing (55.2, 205.0 mtorr), increasing (above 305.0 mtorr), or with a peak (90.0 mtorr). }
    \label{fig:bigg}
\end{figure}

Fig.~\ref{'ir_curve'} studies how pressure affects the current-resistance relation. Here we study how the existence of magnetic fields can affect the plasma resistance at various pressures. In the experiment, We set the power supply to maximum output power, and keep it unchanged throughout the measurement. The only thing that changes during the experiment is the strength of the magnetic field in the negative glow region. The measurements are  shown in Fig.~\ref{fig:bigg}.\\

Interestingly, we realized that the relationship between current, resistance, and magnetic field varies with pressure. From the images we took by CCD, we see that the magnetic field is capable of changing the length of the negative glow region. It could be extended to the cathode, or reduced to a shorter length, as shown in Fig.~\ref{fig:typical_lens}(b).\\

\begin{figure}[H]
    \centering
    \begin{subfigure}{0.45\textwidth}
        \centering
        \includegraphics[width=\linewidth]{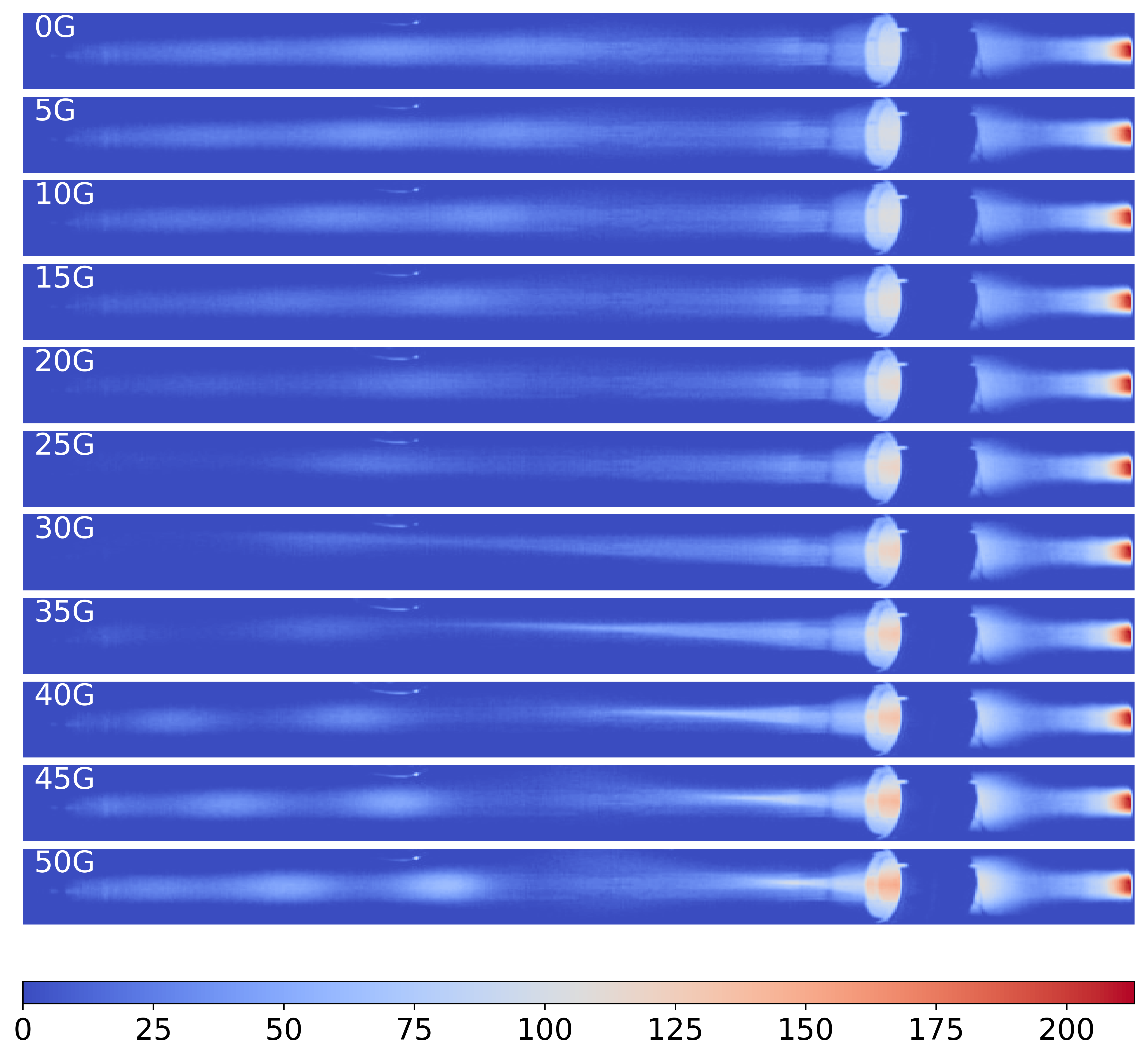}
        \caption{\textbf{Plasma structure at 55.2 mtorr. Resistance decreases with magnetic field.}}
        \label{fig:image1}
    \end{subfigure}
    
    \vspace{0.5cm} 

    \begin{subfigure}{0.45\textwidth}
        \centering
        \includegraphics[width=\linewidth]{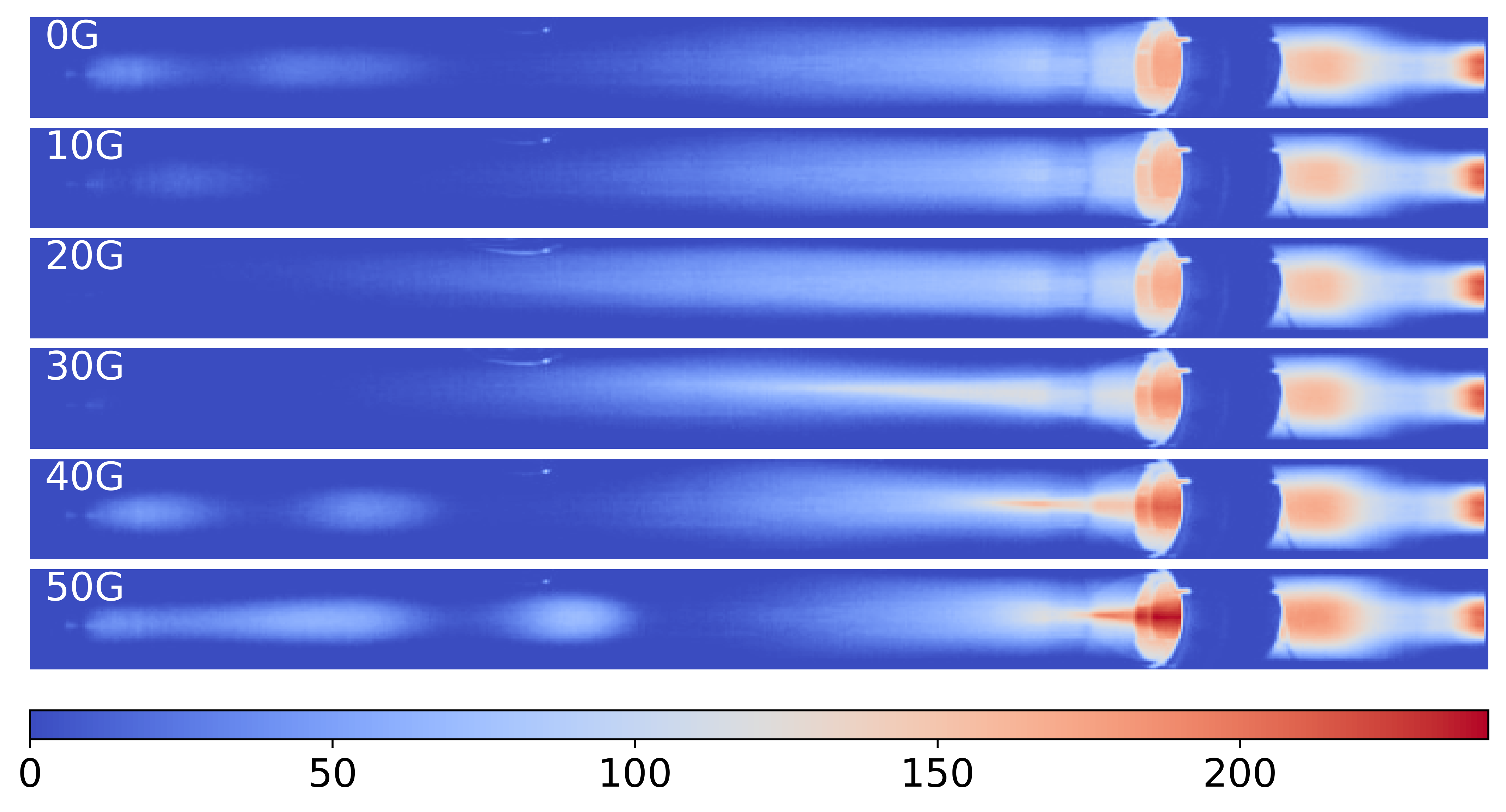}
        \caption{\textbf{Plasma structure at 90.0 mtorr. Resistance peaks at 20 G.}}
        \label{fig:image2}
    \end{subfigure}
    
    \vspace{0.5cm} 

    \begin{subfigure}{0.45\textwidth}
        \centering
        \includegraphics[width=\linewidth]{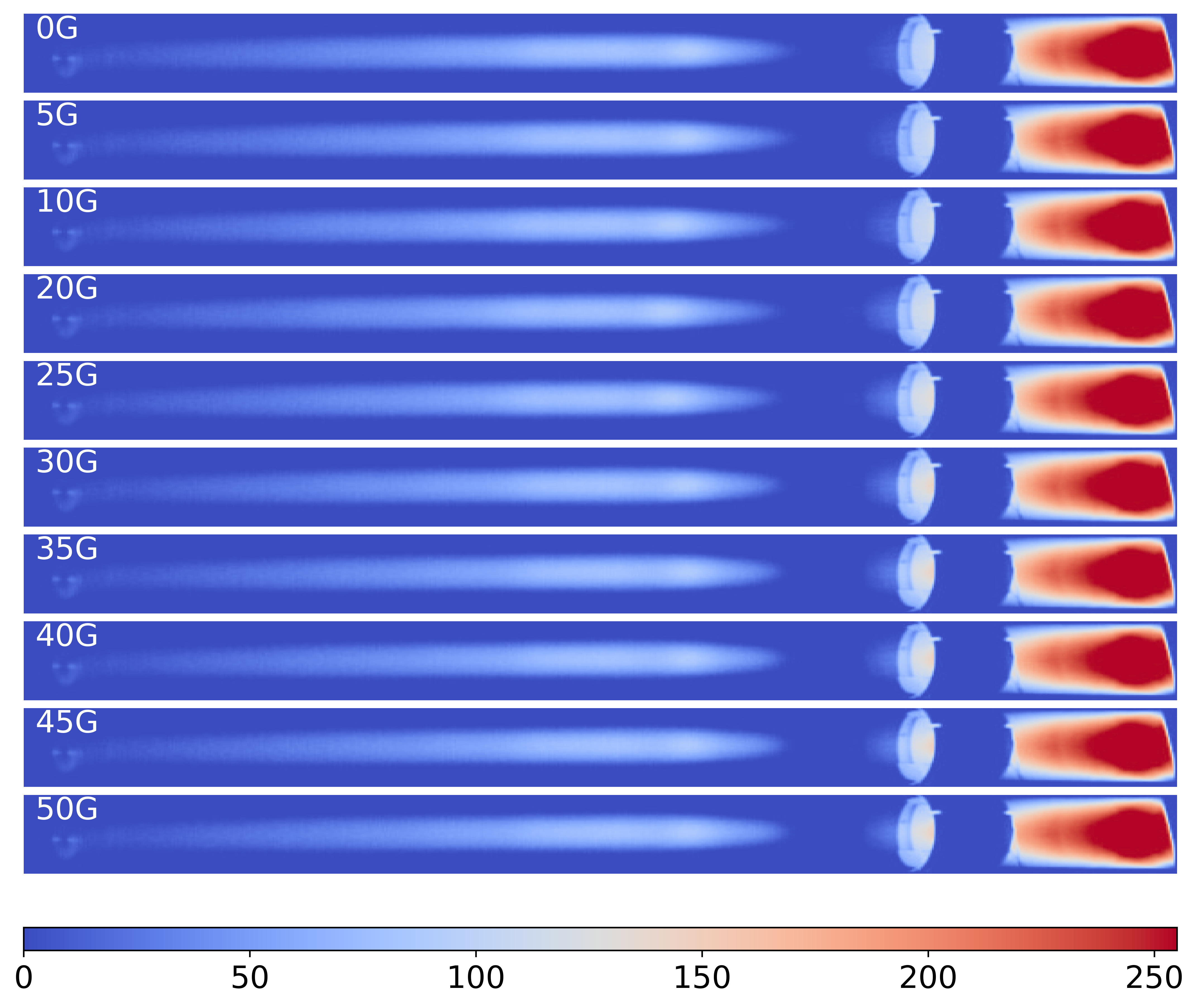}
        \caption{\textbf{Plasma structure at 305.0 mtorr. Resistance increases with magnetic field.}}
        \label{fig:image3}
    \end{subfigure}
    
    \caption{\textbf{Plasma Structure at Various Pressures and Magnetic fields.} The magnetic field is from 0 to 50 G. For (a), as magnetic field increases, the plasma evolves from pure negative glow to some striations. For (b), initially there are few striations, than completely disappears at 20 G. The positive column reappears at higher magnetic field. For (c), the chamber is almost pure positive column, and there is no obvious change as the magnetic field increases.}
    \label{fig:3x1layout}
\end{figure}
We study the three cases: Resistance increase/first-increase-than-decrease/decrease with increasing magnetic field at 55.2/90.0/305.0 mtorr. In Fig.~\ref{fig:3x1layout}. We observed that the resistance seems to go with the length of the negative glow. If the negative glow region is longer, the resistance is higher. In contrast, if the positive column is longer (i.e., more standing striations), then the resistance is lower.
As shown in Fig.~\ref{fig:3x1layout}. For 55.2 mtorr, the negative glow dominates the chamber when no magnetic field is applied. After turning on the magnetic field, the electrons become focused and then spread out after the focal point, causing them to hit the chamber wall and lose energy that would otherwise excite neutral particles. This leads to the formation of a clearer Faraday dark space and a positive column, which lowers the discharge resistance. Therefore, at 55.2 mtorr, the resistance decreases with increasing magnetic field.

For 90 mtorr, striations already appear near the anode even without an applied field. When the magnetic field is increased to around 20 G, the electrons are focused strongly enough that they do not lose energy through wall collisions, allowing the negative glow to extend through the entire chamber. During this process, the resistance increases. Beyond 20 G, however, the resistance begins to decrease again as the positive column reappears, similar to the behavior seen at 55.2 mtorr.

For 305.0 mtorr, although no significant changes in the visible plasma structure are observed, we suspect that the magnetic lens effect still lengthens the positive column, which reduces the resistance.

The reason that the more tightly focused the negative glow is, the lower the resistance, is because with the same electron number passing through the chamber, it only needs to pass through a smaller region, which reduces the number of neutral gas atom that the electron beam needs to ionize to get to the anode.

\subsection{Electron Trajectory Simulation}\label{sim}
To understand the focusing mechanism, and estimate the drift velocity of the electrons, we wrote a Python-based electron trajectory simulation. The simulation incorporates the Runge-Kutta 4 algorithm, which computes Newton's equation of motion with a higher convergence rate than Newton's method. The equation of motion of the electrons is given by the Lorentz force:

\begin{equation}\label{force}
    F(r,z) = -e(v\times B(r,z)) = m_ea
\end{equation}

Where $F(r,z)$ is the force on the electron by the magnetic field at the $(r,z)$ position in cylindrical coordinates, $e$ is the magnitude of electron charge, $B(r,z)$ is the magnetic field, $m_e$ is the electron mass, $a$ is the acceleration. $B(r,z)$ is computed numerically by the Biot-Savart law:

\begin{equation}
    B(r,z) = \frac{\mu_0}{4\pi}\int_{magnetic lens}\frac{K(x,y,z)\times r}{r^3} dA =  \frac{\mu_0}{4\pi}\sum_{z}\sum_{ \theta}\frac{K(\theta,z)\times r(\theta, z)}{r^3}
\end{equation}

\begin{figure}[H]
    \centering
    \includegraphics[width=\textwidth]{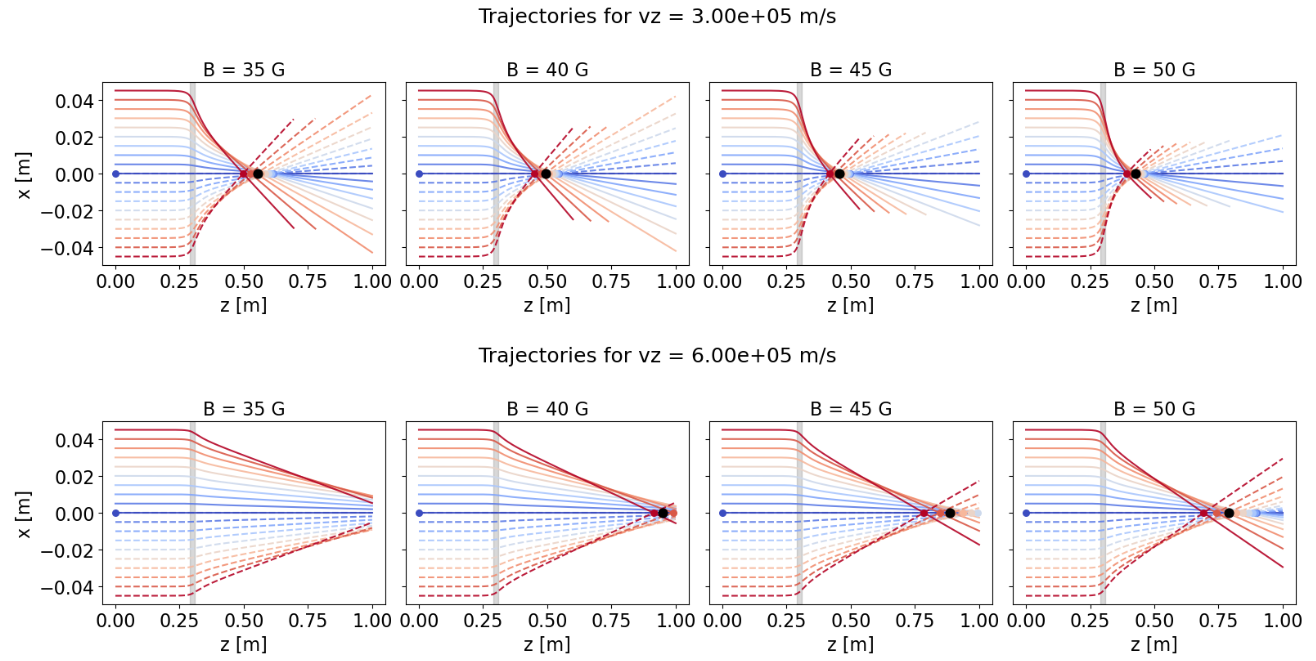}
    \caption{\textbf{Electron trajectory simulation of drift velocity $3\times10^5$ and $3\times10^6$ m/s} The cathode is located at z = 0 m, and the anode at z = 1 m. The various colors correspond to trajectories with various initial positions, from x = 0 m to x = 0.045 m, and the points with same color indicates the focal point of that trajectory, where the black point is the weighted focal point. The shaded region at z = 0.3 m corresponds to the magnetic lens. One can observe the stronger the magnetic field is, or the slower the electron beam is, the shorter the focal length, an agreement with equation \ref{focus}.}
    \label{fig:simulation}
\end{figure}

In the simulation, we assume the electrons emitted by the cathode have some common drift velocity $v_z$ that is independent of the position of emission. We then set the initial position of the electrons with 0.5 cm separation along the x axis. This simplification is due to the rotational symmetry along the z axis of the system. The electron cyclotron frequency at 50 G is 140 MHz, so we set the timestep to be 1 ns and capable of resolving the cyclotron motion. As the electrons move towards the anode, they enter the magnetic lens region at z = 30 cm (same as the experiment), and the trajectory is bent. The dynamics within this region is computed by equation \ref{force}. Examples of the simulation result are shown in Fig.~\ref{fig:simulation}. For each run, we vary three initial conditions: the magnetic field in the magnetic lens, the drift velocity of the electrons, and the initial position of emission. We vary these parameters to compare with the experimental results.\\

To determine the focal length, we interpolate each simulated trajectory to find the position where it crosses the $z$ axis ($x = 0$). For a given magnetic field and drift velocity, trajectories starting from different initial emission positions generally intersect the $z$ axis at different points, yielding a distribution of focal positions. The effective focal point is then obtained as a weighted average of these individual focal positions. The weight assigned to each trajectory is proportional to its initial emission radius, reflecting the assumption of a uniformly emitting circular cathode, for which the available emitting area at radius $r$ scales as $2\pi r$.

\subsection{Obtaining the Electron Drift Velocity}
\begin{figure}[H]
    \centering
    \includegraphics[width=0.6\columnwidth]{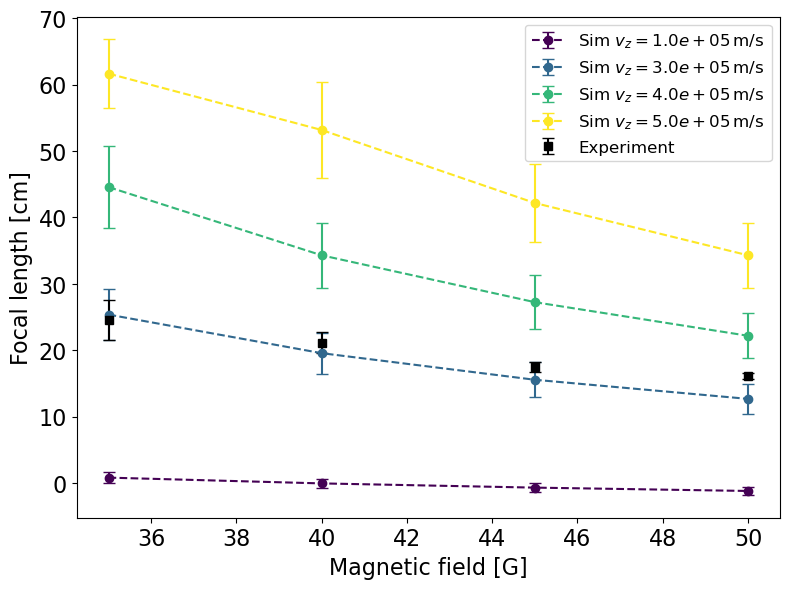}
    \caption{\textbf{Experimentally measured and simulated focal length at various magnetic fields.} The colored points correspond to the simulated focal length at various initial velocities, while the black squares are the experimental results. The experimental result roughly correspond to a simulated electron beam with drift velocity of $3\times10^5$ m/s.}
    \label{sim_com}
\end{figure}

We simulated the electron trajectories with drift velocities ranging from $10^5$ m/s to $10^7$ m/s. The comparison between the experimentally measured focal point and the simulation (Fig.~\ref{sim_com}) shows a drift velocity of  $3\times10^5$ m/s, which corresponds to 0.26 eV of kinetic energy. We do not include a statistical error bar here since we're directly comparing the simulation result to experimental value, not by fitting. To assess whether this drift velocity is physically reasonable, we compare it
with an independent estimate obtained using a separate method.
This is by using the charge conservation equation. The current going through the plasma is related to its current density $J=\frac{I_p}{A}$ and the current density is related to the drift velocity by $J = n_ev_d$. Combining these equations give:

\begin{equation}
v_d = \frac{I_p}{A\, n_e\, e}
\end{equation}

From previous sections, we have independently measured these parameters. In section \ref{langmuir}, we used Langmuir probes to measure the electron density, giving a value of $5.4(\pm 3.2)\times10^7/cm^3$. The helium pressure for the present experiment is ten time lower than that of Fig.\ref{fig:radialIV}. Assuming the ionization rate of the DC discharge plasma remains unchanged when the pressure is reduced, the plasma density should scale proportionally with the neutral density. Under this assumption, we can estimate the electron density in the current conditions by reducing the measured density by the same factor of ten ($5.4(\pm 3.2)\times10^6/cm^3$). In section \ref{IR}, we used a micro-amp meter to measure the plasma current, giving an averaged value of $0.6 (\pm 0.2)~mA$. Assuming the effective area of the plasma has diameter 9 cm, this gives rise to a drift velocity of:

\begin{equation}\label{}
v_d = \frac{0.6 \pm 0.2 mA}{(4cm)^2\pi\ \times5.4(\pm 3.2)\times10^6/cm^3\times 1.602\times10^{-19}}=1.4(\pm0.95)\times10^5~m/s
\end{equation}

Despite the magnetic lens measurement of the electron drift velocity ($3\times10^5$ m/s) exceeds the upper limit of the calculated value $1.4(\pm0.95)\times10^5$ m/s, the correspondence in magnitude suggests that our simplified electron-focusing model reflects the drift velocity of electrons emitted from the cathode.

\section{Conclusions and Future Directions}\label{chap:conclusion}
\subsection{Conclusion}
In this paper, we developed and characterized a DC glow discharge plasma experiment intended for use in an undergraduate teaching laboratory. With this platform, we performed plasma diagnostics, including Paschen breakdown characterization, plasma I–V curve measurements, Langmuir probe analysis of electron temperature and density, and Boltzmann plot spectroscopy. We also investigated plasma dynamics and demonstrated magnetic lensing effects in the plasma. Our results show that this platform is flexible and capable of revealing a wide range of plasma physics phenomena, allowing students to explore fundamental processes such as ionization, atomic excitation, and charged-particle motion. This experiment is now ready for implementation in the undergraduate physics curriculum, where it can serve as a platform for future student-led investigations.

\subsection{Future Work}
Based on this work, several additional topics emerge that could be explored in future experiments with this apparatus:\\
\textbf{1. Microwave interferometry:} This work includes measuring plasma temperature and plasma density by Langmuir probes (section \ref{chap:diag}). Other ways of measuring plasma density is through microwave interferometry. The dispersion relation of cold, unmagnetized plasma is given by:
\begin{equation}
    \omega^2 = \omega_p^2 + c^2 k^2
\end{equation}
Where $\omega_p$ is the electron plasma frequency, given by:
\begin{equation}
    \omega_p = \sqrt{\frac{n_e e^2}{\varepsilon_0 m_e}}
\end{equation}
Where $n_e$ as the electron density, $m_e$ as the electron mass. By performing a microwave interferometry and measuring the phase shift after propagating through the plasma medium, the plasma density can be directly inferred.\\
This measurement would involve setting up microwave horn antennas, microwave mixers, and a microwave source with known frequency. With a variable frequency microwave source and broad band antennas, one would be able to measure the dispersion relation directly.\\
\textbf{2. Double Langmuir Probe:}
The main technical challenge of using a single Langmuir probe for plasma diagnostics is the large variation of the plasma potential. To perform accurate measurements, the probe must be biased to (or swept around) the plasma potential, which in turn requires an additional power supply capable of providing a voltage range comparable to that of the DC supply driving the discharge. In a laboratory equipped with only a single high voltage power supply, this is not feasible.\\
A practical alternative is the double Langmuir probe configuration, in which two closely spaced probes share a common plasma potential. By biasing one probe relative to the other and measuring the resulting current drawn from the plasma, one can still extract key plasma parameters. This arrangement only requires a moderate (50 V) floating power supply referenced to the plasma potential, making it much more suitable for labs that have only one main high voltage supply.

\subsection{Acknowledgment}

We gratefully acknowledge Chang-Jen Hsu for the funding support that enabled the initiation of this work, and for his valuable technical assistance, discussions, and initial provision of laboratory space, all of which were essential to the realization of this project. We further wish to thank Prof. Li-Min Wang and Prof. Yuan-Huei Chang for their assistance in machining components and for lending us vacuum equipment. Finally, we are sincerely thankful to Dr. Yung-Kun Liu for his thoughtful discussions and allowing us to share his lab space.

\bibliographystyle{unsrt}
\bibliography{ref}

\end{document}